\documentclass{article}

\usepackage{arxiv}

\usepackage[utf8]{inputenc} 
\usepackage[T1]{fontenc}    
\usepackage{hyperref}       
\usepackage{url}            
\usepackage{booktabs}       
\usepackage{tabularx}       
\usepackage{ragged2e}
\usepackage{array}
\usepackage{multirow}        
\usepackage[table]{xcolor}
\usepackage{arydshln}       
\ADLinactivate             
\usepackage{amsfonts}       
\usepackage{amsmath}        
\usepackage{nicefrac}       
\usepackage{microtype}      
\usepackage{cleveref}       
\usepackage{lipsum}         
\usepackage{graphicx}
\usepackage{rotating}       
\usepackage{subcaption}
\usepackage{natbib}
\usepackage{doi}
\usepackage{listings}
\lstdefinestyle{tasktemplate}{
  basicstyle=\ttfamily\footnotesize,
  columns=fullflexible,
  keepspaces=true,
  breaklines=true,
  breakatwhitespace=true,
  showstringspaces=false,
  frame=single,
  rulecolor=\color{black!25},
  framesep=5pt,
  xleftmargin=6pt,
  xrightmargin=6pt,
  aboveskip=8pt,
  belowskip=8pt,
  captionpos=t
}

\title{Scientific Coding with Large Language Models: Evidence from Finite Element Analysis}

\newif\ifuniqueAffiliation
\uniqueAffiliationtrue

\ifuniqueAffiliation 
\author{ \href{https://orcid.org/0009-0008-3262-7533?}{\includegraphics[scale=0.06]{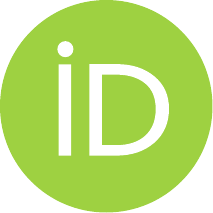}\hspace{1mm}Erfan Hamdi} \\
	Department of Mechanical Engineering\\
	Boston University\\
	Boston, MA 02215 \\
	\texttt{erfan@bu.edu} \\
	\And
	Joel Shor \\
	AI Bio Design, Allen Institute \& Move37 Labs\\
	\texttt{joel.shor@alleninstitute.org} \\
	\And
	\href{https://orcid.org/0000-0001-8099-3468}{\includegraphics[scale=0.06]{orcid.pdf}\hspace{1mm}Emma Lejeune} \\
	Department of Mechanical Engineering\\
 	Division of Systems Engineering\\
	Boston University\\
	Boston, MA 02215 \\
	\texttt{elejeune@bu.edu} \\
}
\fi

\renewcommand{\shorttitle}{Scientific Coding with Large Language Models}

\hypersetup{
pdftitle={Scientific Coding with Large Language Models},
pdfsubject={},
pdfauthor={Erfan Hamdi, Joel Shor, Emma Lejeune},
pdfkeywords={Scientific computing, Large language models, Coding agents, Self-verification, Failure taxonomy, Finite element method, Computational mechanics},
}


\begin{document}
\raggedbottom
\maketitle

\begin{abstract}
Large language models (LLMs) offer a potential route to automating scientific programming, but their usefulness depends on consistently producing numerically correct code. We extend FEM-Bench to 17 finite element tasks involving nonlinear and history-dependent mechanics and structural dynamics, and evaluate 15 models on implementation and test generation without execution feedback. The best model succeeds in 87.1\% of single-pass implementation trials, but rankings differ between tracks. The leading implementer achieves 49.9\% joint test success, while the leading tester reaches 82.6\%. For six models, we compare single-pass generation with two agentic harnesses that allow code execution and revision. Agentic execution solves some tasks that never succeeded in single-pass trials and improves consistency on others, but gains vary across models and harnesses, and regressions also occur. Claude Opus 5 with SimpleAgent reaches 96.5\% Pass@3, the probability of at least one success in three attempts, but 77.1\% Pass$^3$, the probability that all three succeed. No model-harness combination exceeds 78.8\% Pass$^3$. To understand why agents submit incorrect code after passing their own checks, we analyze 249 failed agentic trajectories and identify six categories of self-verification failure. The most common are inadequate test coverage (73.1\%) and reliance on generic consistency checks (70.3\%), and in 28.5\% of cases the tests share an error with the implementation. Motivated by these findings, an exploratory configuration with separate coder and tester agents achieves the highest observed scores, with 97.6\% Pass@3 and 88.8\% Pass$^3$. These results identify consistency across attempts and independent verification as the main obstacles to trusting LLM-generated scientific code.
\end{abstract}


\section{Introduction}
Large language models (LLMs) are increasingly being used to solve mathematical problems and perform computational tasks in scientific research~\citep{lewkowycz2022solving,chen2025scienceagentbench}. Their ability to generate code from natural-language specifications has been demonstrated on tasks ranging from self-contained functions to programs that combine multiple software libraries~\citep{chen2021evaluating,zhuo2025bigcodebench}. For scientists and engineers, these capabilities offer a potential route to automating the implementation of computational models and the analysis of their outputs. Agentic harnesses extend code generation by connecting models to tools for executing programs, inspecting results, and revising their implementations~\citep{wang2024openhands}. This interaction allows models to use execution feedback during development and may support more autonomous scientific programming. A central question is how reliably these capabilities extend to scientific computations that must satisfy both numerical and physical requirements.

These numerical and physical requirements are central to computations governed by partial differential equations (PDEs). Many problems in solid mechanics, fluid flow, and heat transfer are described by PDEs, which relate physical fields to their spatial and temporal derivatives \citep{courant1989methods,strauss2008partial}. Classical approaches such as finite differences, finite volumes, and finite elements approximate their solutions through discretization \citep{leveque2007finite,leveque2002finite,hughes2000finite}. In finite element analysis, implementing these approximations requires consistent treatment of element formulations, material behavior, boundary conditions, and solution algorithms~\citep{hughes2000finite}. Numerical accuracy depends on discretization and solver convergence, while floating-point arithmetic can introduce rounding errors~\citep{goldberg1991floating}. Verification must therefore distinguish acceptable numerical discrepancies from implementation errors and use tolerances appropriate to the quantities being checked. Analytical solutions, convergence studies, and checks of relevant physical properties provide complementary ways to assess a computation~\citep{oberkampf2002verification}. These checks must also cover the geometries, loading conditions, and material regimes specified by the task. A program can execute successfully and produce plausible outputs while still implementing the wrong computation. Together, these demands make finite element analysis an informative setting for assessing both scientific code generation and the reasoning used to verify its results.

Here, we investigate how reliably LLMs can implement specified finite element computations and generate tests that distinguish correct from incorrect code. We also examine how execution and revision affect performance and how agents assess their own implementations. We use tasks with explicit mathematical requirements and independent numerical checks to probe scientific reasoning through implementation and verification behavior. Our broader motivation is to assess capabilities that LLMs would need to serve as co-scientists or co-engineers responsible for the computational components of a project. We build on our earlier FEM-Bench study, which evaluated introductory finite element and structural analysis tasks in a single-pass setting without execution feedback~\citep{mohammadzadeh2026fembench}. The present work extends this evaluation to more demanding tasks involving nonlinear mechanics, history-dependent material behavior, and structural dynamics. We compare single-pass generation with agentic workflows that support execution and revision, and analyze verification failures in completed but incorrect agentic submissions.

We evaluate 15 models on 17 tasks in separate single-pass implementation and test-generation tracks. Generated implementations are compared with reference outputs on prescribed inputs, while generated tests are evaluated by whether they accept reference implementations and reject designated faulty ones. For six of these models, we compare single-pass generation with two agentic settings: SimpleAgent and OpenHands. We assess implementation success, consistency across attempts, and cost. We also explore a SimpleAgent configuration with separate coder and tester agents. Each setting uses five trials per task and model. To study self-verification, we analyze the trajectories of 249 completed SimpleAgent and OpenHands trials whose submitted implementations fail independent evaluation. We develop a codebook through manual review and pilot judging. LLM judges use it to assess the agents' checks, interpretation of results, and verification of subsequent revisions.

Our contributions are threefold. First, we extend FEM-Bench to more demanding mechanics tasks and evaluate scientific coding in single-pass and agentic settings. Second, we show strong implementation performance, with the best single-pass model succeeding in 87.1\% of trials. Pass@3 estimates the probability of at least one successful implementation in three trials. Pass$^3$ estimates the probability that all three succeed. For Claude Opus 5, Pass@3 rises from 90.6\% with single-pass generation to 96.5\% with SimpleAgent. Its corresponding Pass$^3$ increases from 65.9\% to 77.1\%. An exploratory coder--tester configuration reaches 97.6\% Pass@3 and 88.8\% Pass$^3$. These are the highest observed scores among the evaluated configurations. Agentic gains vary across models and harnesses, and regressions also occur. Model rankings also differ between implementation and test generation. Third, we identify six categories of self-verification failure in completed, incorrect agentic submissions. These categories describe weaknesses in test coverage, numerical checks, and the handling of test results and code revisions.

Section~\ref{sec:related_work} reviews related scientific coding benchmarks and agent verification studies. Section~\ref{sec:tasks_and_evaluation} presents the tasks, evaluation procedure, and metrics. Section~\ref{sec:experimental_setup} describes the experimental settings and self-verification analysis procedure. Section~\ref{sec:results_discussion} presents the results and discusses their implications and limitations. Section~\ref{sec:conclusion} concludes the paper. The appendices provide detailed results, task specifications, harness settings, and judging procedures.

\section{Related Work}
\label{sec:related_work}
Code-generation benchmarks evaluate both general programming and scientific tasks. HumanEval and MBPP test short, self-contained programs~\citep{chen2021evaluating,austin2021program}. DS-1000 and BigCodeBench test programming with software libraries~\citep{lai2023ds1000,zhuo2025bigcodebench}, while SWE-bench asks models to resolve issues in existing repositories~\citep{jimenez2024swebench}. Scientific benchmarks place coding tasks in research settings. SciCode divides scientist-written research problems into subproblems. Its tests compare generated outputs with numerical references, analytical solutions, and published results~\citep{tian2024scicode}. ScienceAgentBench and MLAgentBench cover scientific data analysis and machine learning experiments~\citep{chen2025scienceagentbench,huang2023mlagentbench}. Our focus within scientific programming is on implementing finite element computations and checking their numerical correctness.

Closely related work examines how models generate PDE solvers and use numerical libraries. CodePDE asks models to implement solvers from equations, initial and boundary conditions, and specified function interfaces. It measures numerical error, convergence, and runtime, and studies debugging and refinement~\citep{li2026codepde}. PDEAgent-Bench focuses on solvers built with finite element libraries such as DOLFINx, Firedrake, and deal.II. It checks whether code runs, meets numerical accuracy requirements, and finishes within a runtime limit~\citep{hang2026pdeagentbench}. FEABench evaluates LLMs and agents on solving physics and engineering problems through the COMSOL API~\citep{mudur2025feabench}. FEM-Bench evaluates mechanics functions and complete analyses, varying how many helper functions are supplied. It also tests whether model-written tests accept correct implementations and reject known incorrect ones~\citep{mohammadzadeh2026fembench}. We extend FEM-Bench to history-dependent and nonlinear mechanics and structural dynamics. We compare implementation success and consistency across repeated trials in single-pass and agentic settings.

When models can execute and revise code, the evaluation can also examine how they check their own work. Self-Debugging uses execution results and code explanations to help models revise their own programs~\citep{chen2024selfdebugging}. \citet{olausson2024selfrepair} find that self-repair gains are often modest once repair costs are included. They also report higher repair success with feedback from stronger models or human programmers. Studies of agent failures use taxonomies that cover reasoning, execution, and coordination errors, as well as incomplete or incorrect verification~\citep{cemri2025fail,deshpande2025trail}. In scientific research, \citet{kim2026unknown} describe agents accepting limited checks as evidence that broader task requirements have been met. We examine self-verification during finite element code development. Our analysis covers completed agent runs whose final code fails independent evaluation. We trace what agents check, how they interpret the results, and whether they verify the code they finally submit.

\section{Tasks and Evaluation}
\label{sec:tasks_and_evaluation}
We evaluate scientific coding through two complementary capabilities: (1) implementing finite element computations and (2) generating tests that distinguish correct from incorrect implementations. Section~\ref{sec:tasks} introduces the task suite and the specifications and supporting code provided to the models. Section~\ref{sec:evaluation_procedure} describes how generated implementations and tests are evaluated. Section~\ref{sec:evaluation_metrics} defines the metrics used to assess implementation correctness, test effectiveness, and consistency across repeated attempts.

\subsection{Tasks}
\label{sec:tasks}
We develop 17 finite element tasks designed to present substantially greater complexity than the introductory task suite evaluated in the first FEM-Bench study~\citep{mohammadzadeh2026fembench}. The tasks include history-dependent elastoplastic updates, nonlinear material and interface behavior, plate bending, three-dimensional continuum mechanics, and time-dependent structural response. These computations require handling constitutive state, coordinating multiple stages of a finite element solution, and/or maintaining consistent updates across load increments and time steps. The suite also includes tasks for mesh generation and plate-element shape functions. Table~\ref{tab:fem_bench_tasks} summarizes the tasks and their helper-function variants.

Each task is presented through a Python function signature and a detailed docstring specifying the required computation. The docstring describes the physical problem and numerical formulation and defines the inputs, outputs, and required behavior of the function. Depending on the task, it specifies array dimensions, connectivity and degree-of-freedom ordering, material parameters, loading and boundary conditions, solver settings, and conditions under which the function should raise an error.

Supporting information includes permitted imports and, where applicable, supplied helper functions. Several tasks have multiple variants with different amounts of helper functions provided, changing which parts of the code the model must implement. For test generation, the LLM receives the requested test function's specification, including its name and description. Appendix~\ref{app:task_prompt_templates} presents the task structure and an example from the 3D dynamics task. The implementation and test-generation prompt templates are provided in the study repository.

Task names are identified with the H/T notation introduced in FEM-Bench~\citep{mohammadzadeh2026fembench}. \textbf{H} denotes the number of helper functions used by the reference implementation, for example H0 indicates a reference implementation without helpers, whereas H5 indicates a reference implementation that uses five.

\textbf{T} describes the availability of helper functions in the implementation prompt. T0 applies when the reference implementation uses no helper functions and none are supplied. T1 supplies all helper functions, T2 supplies a subset, and T3 withholds all helper functions. 
Tasks in the same family but at different tiers keep the same target function signature. The difference is that, for tasks with helper functions, the prompt provides complete implementations only for the available helpers.
Thus, H is the helper function count, while T specifies how much of that supporting code is available to the model. For example, the Kirchhoff plate tasks are all listed as H5 where the T1 variant supplies five helpers, the T2 variant supplies only the shape-function helper, and the T3 variant supplies none.

\begin{table}[!htbp]
\centering
\caption{\textbf{Summary of the 17 tasks evaluated in this study.} Tasks are grouped by domain. H denotes the reference-helper count, and T the helper-availability tier. Section~\ref{sec:tasks} defines these annotations.}
\label{tab:fem_bench_tasks}
\renewcommand{\arraystretch}{1.15}
\setlength{\tabcolsep}{4pt}
\begin{tabularx}{\linewidth}{@{}l >{\raggedright\arraybackslash}p{0.27\linewidth} >{\raggedright\arraybackslash}X c c@{}}
\toprule
\textbf{Domain} & \textbf{Task Name} & \textbf{Description} & \textbf{H} & \textbf{T} \\
\midrule
\textit{FEM 1D} & isotropic hardening & Elastoplastic stress update with linear isotropic hardening & 0 & 0 \\
\textit{FEM 1D} & kinematic hardening & Elastoplastic stress and back-stress update with linear kinematic hardening & 0 & 0 \\
\textit{FEM 1D} & linear elastic dynamics & Elastic bar response using quadratic elements and explicit Newmark integration & 1 & 1 \\
\midrule
\textit{FEM 2D} & circular mesh & Triangular mesh on a circular domain using prescribed connectivity & 0 & 0 \\
\textit{FEM 2D} & circular mesh (Delaunay) & Triangular mesh on a circular domain using Delaunay triangulation & 0 & 0 \\
\textit{FEM 2D} & Specht shape functions and derivatives & Shape functions and first and second derivatives for a Specht triangular element & 0 & 0 \\
\textit{FEM 2D} & Kirchhoff plate bending & Solve thin-plate bending using Specht triangular elements & 5 & 1 \\
\textit{FEM 2D} & Kirchhoff plate bending & Solve thin-plate bending using Specht triangular elements & 5 & 2 \\
\textit{FEM 2D} & Kirchhoff plate bending & Solve thin-plate bending using Specht triangular elements & 5 & 3 \\
\textit{FEM 2D} & hyperelasticity & Solve plane-strain neo-Hookean elasticity using quadrilateral elements & 10 & 1 \\
\textit{FEM 2D} & hyperelasticity & Solve plane-strain neo-Hookean elasticity using quadrilateral elements & 10 & 3 \\
\textit{FEM 2D} & cohesive zone & Solve elasticity with nonlinear cohesive interfaces using Newton--Raphson iteration & 13 & 1 \\
\textit{FEM 2D} & cohesive zone & Solve elasticity with nonlinear cohesive interfaces using Newton--Raphson iteration & 13 & 2 \\
\textit{FEM 2D} & cohesive zone & Solve elasticity with nonlinear cohesive interfaces using Newton--Raphson iteration & 13 & 3 \\
\midrule
\textit{FEM 3D} & hex8 linear elasticity & Displacements, strains, stresses, and reactions using eight-node hexahedral elements & 8 & 1 \\
\textit{FEM 3D} & linear elastic dynamics & Structural dynamics using hexahedral elements and Newmark integration & 5 & 1 \\
\textit{FEM 3D} & linear elastic dynamics & Structural dynamics using hexahedral elements and Newmark integration & 5 & 3 \\
\bottomrule
\end{tabularx}
\end{table}

\subsection{Evaluation procedure}
\label{sec:evaluation_procedure}
We evaluate two capabilities: implementing finite element computations from a detailed specification, and generating tests that distinguish correct from incorrect implementations. In the implementation track, the requested output is a Python function satisfying the prescribed signature and computational requirements. In the test-generation track, the requested outputs are test functions addressing specified aspects of the target function's behavior.

Generated implementations and tests are evaluated using the FEM-Bench evaluation framework~\citep{mohammadzadeh2026fembench}. LLM generated codes for the implementation track are executed on a set of predefined verification inputs, and their outputs are compared with those of reference implementations. Generated tests are evaluated against both the reference implementation and designated incorrect implementations labeled as ``Expected Failure'' cases. A successful test must accept the reference implementation and reject the expected failure cases in the task.

\subsection{Evaluation metrics}
\label{sec:evaluation_metrics}
For each model and evaluated configuration, we use $M=17$ tasks and $n=5$ trials per task. A trial is one complete run. In agentic settings, it includes all the intermediate steps before the final code submission by the agent. All metrics are reported as percentages, with equal weight assigned to each task. Function correctness, Pass@3, and Pass$^3$ use implementation outcomes. Joint test success rate is used only in the test-generation track.

\paragraph{Function correctness.}
Let $x_{ir}=1$ if the implementation produced in trial $r$ of task $i$ passes the benchmark evaluation, and $x_{ir}=0$ otherwise. Success requires agreement with the reference implementation on all predefined verification inputs under the evaluator's comparison criteria. Runs with no evaluated submission, including those that exhaust the budget, contribute no successes. Let $c_i=\sum_{r=1}^{n}x_{ir}$ be the number of successful trials. Mean function correctness (FC) is:
\begin{equation}
\label{eq:function_correctness}
\mathrm{FC}
= \frac{1}{M}\sum_{i=1}^{M}\frac{c_i}{n}
= \widehat{\mathrm{Pass@1}}.
\end{equation}
Function correctness equals empirical Pass@1. It estimates the probability of success in a single trial, averaged over tasks.

\paragraph{Joint test success rate.}
A generated test succeeds only if it accepts the reference implementation and rejects expected failure cases associated with the task. This criterion rewards both acceptance of correct behavior and detection of faults. Each task requires at least one test case. For a trial with completed test evaluation, let $m_{ir}\geq0$ be the number of test cases generated by the LLM. For test case $j$ in trial $r$ of task $i$, we define
\begin{equation}
\label{eq:joint_test_indicator}
z_{irj}=
\begin{cases}
1, & \text{accepts the reference AND rejects all assigned expected failures,}\\[3pt]
0, & \text{otherwise.}
\end{cases}
\end{equation}
We compute the per-trial rate $s_{ir}$ and its average as
\begin{equation}
\label{eq:joint_test_success}
s_{ir}=\frac{1}{m_{ir}}\sum_{j=1}^{m_{ir}}z_{irj},
\qquad
\mathrm{JTS}=\frac{1}{M}\sum_{i=1}^{M}\frac{1}{n}\sum_{r=1}^{n}s_{ir}.
\end{equation}
When $m_{ir}=0$, we set $s_{ir}=0$. This metric gives partial credit when some test cases succeed, and does not require the entire generated test suite to succeed. Averaging per-trial rates ensures that tasks with more test cases do not receive greater weight.

\paragraph{Pass@3.}
Pass@3 measures the probability of obtaining at least one successful implementation in three independent trials, following the Pass@$k$ criterion used in program synthesis~\citep{kulal2019spoc}. We use the unbiased estimator of \citet{chen2021evaluating} with $k=3$:
\begin{equation}
\label{eq:pass_at_3}
\widehat{\mathrm{Pass@3}}
=\frac{1}{M}\sum_{i=1}^{M}
\left[1-\frac{\binom{n-c_i}{3}}{\binom{n}{3}}\right].
\end{equation}
Here, $c_i$ is the number of successful trials among the $n$ trials for task $i$. The estimator considers all groups of three trials among the five observed trials for each task. Pass@3 captures whether repeated attempts can yield a correct implementation. It does not measure whether the model can identify which implementation is correct.

\paragraph{Pass$^3$.}
Pass$^3$ measures the probability that all three independent trials succeed. Introduced in $\tau$-bench to evaluate reliability across repeated attempts~\citep{yao2024taubench}, Pass$^k$ complements Pass@$k$ by requiring success in every trial. Its estimator for $k=3$ is
\begin{equation}
\label{eq:pass_power_3}
\widehat{\mathrm{Pass}^{3}}
=\frac{1}{M}\sum_{i=1}^{M}
\frac{\binom{c_i}{3}}{\binom{n}{3}}.
\end{equation}
For scientific programming, producing a correct implementation occasionally is insufficient. Numerical errors in generated code can be difficult to detect. Pass$^3$ therefore measures consistency under the same task specification and configuration.

For example, three successful implementations in five trials give a task-level function correctness rate of $60\%$, an estimated Pass@3 of $100\%$, and an estimated Pass$^3$ of $10\%$. This performance can therefore indicate a high chance of finding a successful implementation within three attempts but a low chance that all three attempts succeed.

\section{Experimental Setup}
\label{sec:experimental_setup}

We conduct our experiments on the 17 tasks summarized in Table~\ref{tab:fem_bench_tasks}, using 15 models and five trials per task. All 15 models listed in Table~\ref{tab:models} are evaluated in the single-pass setting through separate implementation and test-generation tracks. Six of these models are also evaluated in the agentic setting with the SimpleAgent and OpenHands harnesses. All agentic settings are evaluated only on the implementation track. Table~\ref{tab:experimental_configurations} summarizes the experimental configurations.

\begin{table}[!htbp]
\centering
\caption{\textbf{Models and single-pass generation settings.} All 15 models were evaluated in the implementation and test-generation tracks. Provider identifies the API service used in the retained runs. Developer identifies the model developer. Output caps are the recorded token limits per call in the single-pass setting. Reasoning denotes the configured setting. Prices are uncached input/output rates in USD per million tokens. Agentic output caps are given separately in Table~\ref{tab:experimental_configurations}.}
\label{tab:models}
\footnotesize
\renewcommand{\arraystretch}{1.2}
\setlength{\tabcolsep}{3pt}
\begin{tabular}{@{}lllllrr@{}}
\toprule
\textbf{Model} & \textbf{Developer} & \textbf{Access} & \textbf{Provider} & \textbf{Reasoning} & \shortstack{\textbf{Output}\\\textbf{cap}} & \shortstack{\textbf{Price}\\\textbf{input/output}} \\
\midrule
Claude Fable 5.1 & Anthropic & Proprietary & AWS Bedrock & Adaptive (high) & 128,000 & 10.00 / 50.00 \\
Claude Opus 5 & Anthropic & Proprietary & AWS Bedrock & Adaptive (high) & 128,000 & 5.00 / 25.00 \\
Claude Sonnet 5 & Anthropic & Proprietary & AWS Bedrock & Adaptive (high) & 128,000 & 2.00 / 10.00 \\
DeepSeek V4 Flash 0731 & DeepSeek & Open weights & Together AI, OpenRouter\textsuperscript{a} & High & 393,216 & 0.14 / 0.28 \\
DeepSeek V4 Pro 0813 & DeepSeek & Open weights & Together AI & High & 393,216 & 1.32 / 3.96 \\
Gemini 3.1 Pro Preview & Google & Proprietary & Google & High & 65,536 & 2.00 / 12.00 \\
Gemini 3.7 Flash & Google & Proprietary & Google & High & 65,536 & 0.75 / 3.75 \\
Gemini 4 Argon & Google & Proprietary & Google & High & 1,048,576 & 4.00 / 20.00 \\
Gemma 4 31B & Google & Open weights & AWS Bedrock & High & 32,768 & 0.14 / 0.40 \\
GLM-5.2 & Z.ai & Open weights & Together AI & High & 131,072 & 1.40 / 4.40 \\
GPT-5.6 Luna & OpenAI & Proprietary & AWS Bedrock & High & 128,000 & 0.22 / 1.32 \\
GPT-5.6 Sol & OpenAI & Proprietary & AWS Bedrock & High & 128,000 & 5.50 / 33.00 \\
GPT-6 Astra & OpenAI & Proprietary & AWS Bedrock & High & 128,000 & 11.00 / 55.00 \\
Kimi K3 & Moonshot AI & Open weights & Together AI, AWS Bedrock\textsuperscript{b} & High & 131,072 & 3.00 / 15.00 \\
Muse Glimmer 30B & Meta & Open weights & Together AI & High & 90,000 & 0.35 / 1.50 \\
\bottomrule
\end{tabular}
\par\smallskip
\begin{minipage}{\linewidth}
\footnotesize
\textsuperscript{a}DeepSeek Flash uses Together AI for 76 runs per track. After DeepSeek V4 Flash 0731 was deprecated on Together AI, we used OpenRouter/Relace and Cohere hosted models with no quantizatoin for nine implementation and test-generation runs with the same token cap. Token costs were calculated using the rates provided by Together AI.
\textsuperscript{b}Kimi uses Together AI for 42 implementation runs and AWS for 43. For test generation, it uses Together AI for 51 runs and AWS for 34. The model was unquantized on both providers. Token costs were calculated using the rates provided by Together AI.
\end{minipage}
\end{table}

\begin{table}[!htbp]
\centering
\caption{\textbf{Experimental configurations.} Each setting uses the 17 tasks in Table~\ref{tab:fem_bench_tasks}, with five trials per task and model or coder-tester pairing. Single-pass implementation and test generation are separate tracks. The six shared models and Fable/Astra pairings are identified in Table~\ref{tab:task_model_implementation_tests}.}
\label{tab:experimental_configurations}
\small
\renewcommand{\arraystretch}{1.2}
\setlength{\tabcolsep}{4pt}
\begin{tabular}{@{}lllll@{}}
\toprule
\textbf{Configuration} & \textbf{Single-pass} & \textbf{SimpleAgent} & \textbf{OpenHands} & \textbf{coder-tester} \\
\midrule
Models & 15 models & 6 models & 6 models & Fable/Astra pair\\
Tracks & \begin{tabular}[c]{@{}l@{}}Implementation,\\Test generation\end{tabular} & Implementation & Implementation & Implementation \\
\midrule
Tools and actions & No tools & \begin{tabular}[c]{@{}l@{}}Python execution,\\submit\end{tabular} & \begin{tabular}[c]{@{}l@{}}Terminal, file editor,\\browser, planning, finish\end{tabular} & \begin{tabular}[c]{@{}l@{}}Python execution,\\feedback, submit\end{tabular} \\
Execution state & No execution loop & \begin{tabular}[c]{@{}l@{}}Fresh sandbox for\\each execution\end{tabular} & \begin{tabular}[c]{@{}l@{}}Persistent workspace\\files\end{tabular} & \begin{tabular}[c]{@{}l@{}}Fresh sandbox,\\tests retained\end{tabular} \\
Internet access & --- & Disabled & Enabled & Disabled \\
\midrule
Out-token cap per call & Model-specific & 30,000 & 30,000 & 32,000 \\
Total token budget & --- & 2,000,000 & 2,000,000 & 2,000,000 \\
Wall-time limit & --- & 2 hours & 2 hours & 2 hours \\
Iteration limit & One generation & 200 & 200 & \begin{tabular}[c]{@{}l@{}}200 iterations per agent,\\50 candidate versions\end{tabular} \\
\midrule
Submission rule & --- & \begin{tabular}[c]{@{}l@{}}At least one execution\\and latest exit code 0\end{tabular} & Agent finish action & \begin{tabular}[c]{@{}l@{}}Tester approval and\\nonempty test suite\\with all tests passing\end{tabular} \\
\bottomrule
\end{tabular}
\end{table}

\subsection{Single-pass}
\label{sec:single_pass_setup}
The single-pass setting serves as our baseline. In each trial, the model generates code in a single response to the task prompt. It has no access to execution tools or feedback. The generated code is then evaluated, with no opportunity for the model to revise.
The reference implementation and expected failure cases are withheld from the test-generation prompt and are used subsequently to evaluate the generated tests.
Models evaluated in both implementation and test generation tracks use the providers, reasoning settings, and output-token caps listed in Table~\ref{tab:models}.

\subsection{Agentic harnesses}
A coding agent combines an LLM with tools that allow it to act on its environment and use the resulting observations to guide further actions. In our experiments, agents can execute code, inspect execution feedback, and revise their implementations over multiple iterations before submitting a final code. We evaluate two harnesses: SimpleAgent, a minimal harness developed for this study, and OpenHands, a general-purpose coding-agent framework.

\subsubsection{SimpleAgent}
\begin{figure}[h]
\centering
\includegraphics[width=\linewidth]{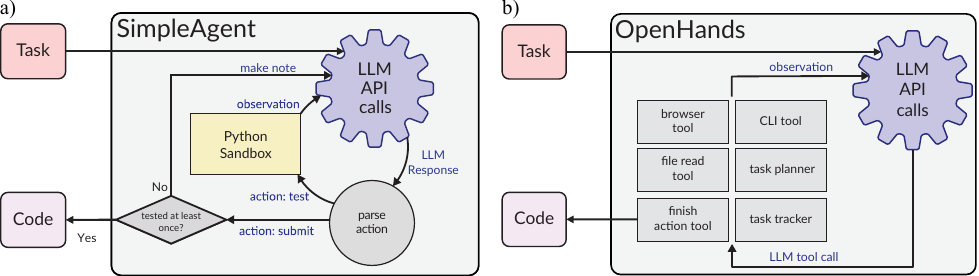}
\caption{\textbf{Overview of agentic harnesses studied.} 
(a) Overview of SimpleAgent. SimpleAgent follows a ReAct-style loop with two actions: Python execution (test) and submission (submit). The Python execution action runs code in a fresh sandbox preloaded with the task's required imports and helper functions. Execution is stateless: variables and files do not persist between executions, while the LLM retains the conversation history and receives execution output and exit codes as observations. Submission is permitted only after at least one execution and only if the most recent execution returned exit code 0.
(b) Overview of the OpenHands coding agent. OpenHands is an open-source framework for coding agents. In this study, the agent had access to tools for web browsing, terminal execution, file editing, task tracking, planning, delegation to subagents, and a finish action to end the session. The agent ran inside a Singularity container configured to isolate it from the local reference implementations while retaining internet access.}
\label{fig:agent_harnesses}
\end{figure}

SimpleAgent is a minimal coding agent that we developed for this study. The harness provides two actions, Python execution and submission, as shown in Figure~\ref{fig:agent_harnesses}(a). The Python execution action runs model-generated code in a fresh sandbox preloaded with the task's required imports and supplied helper functions. Execution is stateless, meaning that the variables and files created during one execution are unavailable in later executions. The model nevertheless has access to the conversation history, including previous actions, execution output, and exit codes. Internet access is disabled.

The prompt instructs the model to implement the requested function, construct and execute tests, and revise its code in response to failures. The harness permits submission only after at least one Python execution and only when the most recent execution returns exit code 0. Otherwise, submission is rejected and the model receives feedback explaining the unmet requirement. An accepted submission ends the trial, and the submitted function is evaluated separately using FEM-Bench. Appendix~\ref{app:harness_details} summarizes the operational limits. The complete prompt templates are provided in the study repository.

\subsubsection{OpenHands}
OpenHands is an open-source framework for coding agents that supports code editing, command-line execution, and web browsing~\citep{wang2024openhands}. In our configuration (Figure~\ref{fig:agent_harnesses}(b)), the agent also had tools for planning, task tracking, and delegation to subagents. Files persisted within the workspace across iterations, allowing the agent to develop and revise implementations and tests over the course of a trial. The agent was instructed to write its final implementation to \texttt{submission.py} and use the finish action to end the session.

Preliminary runs included attempts to locate the reference codes outside the task workspace, motivating isolation of the harness. We ran OpenHands inside a Singularity container configured to isolate the agent from local benchmark reference materials while keeping internet access.
Browser tools were invoked in 30 of the 510 OpenHands runs. External resources were also accessed through terminal commands, including downloads from the public FEM-Bench repository, which contained tasks from the earlier benchmark study in~\citep{mohammadzadeh2026fembench}.
For both SimpleAgent and OpenHands, each trial used a total token budget of 2,000,000, a maximum of 200 iterations, and a wall-time limit of two hours. Model responses were capped at 30,000 output tokens per call. These settings are summarized in Table~\ref{tab:experimental_configurations}. Appendix~\ref{app:harness_details} gives additional operational limits. The task prompt is provided in the study repository.

\subsubsection{Coder-tester extension of SimpleAgent}
\label{sec:coder_tester_setup}
We conduct an exploratory evaluation of separating implementation and verification into two agent roles. This design direction was motivated by the findings in the self-verification error analysis discussed in Section~\ref{sec:self_verification_results}. AgentCoder similarly separates code generation from test design and uses execution feedback for iterative refinement~\citep{huang2024agentcoder}. We extend SimpleAgent with two sub-agents: a coder and a tester. The coder develops the requested function. The tester evaluates candidate implementations and controls final submission. Both agents receive the task specification and maintain separate conversation histories. We evaluate Claude Fable 5.1 and GPT-6 Astra in both role assignments using the same prompts and resource limits. Each pairing is evaluated on the 17 tasks with five trials per task.

The coder can execute its own implementation in a stateless Python sandbox before publishing a candidate to the tester. To encourage regular interaction with the tester, we permit at most three coder executions before each candidate is published. Additional execution requests are rejected until the coder publishes a candidate. The tester receives a callable and the compiled \verb|.pyc| candidate. It does not have access to the Python source. It constructs pytest tests from the task specification and can run them against the compiled candidate to investigate its behavior.

The tester's test suite is kept across iterations and is rerun against each new candidate. The tester can add new tests and rewrite or remove tests from earlier iterations. After investigating the candidate, it sends written feedback to the coder, accompanied by only the names of failing tests. The feedback can describe observed discrepancies, explain suspected causes, and suggest corrections. The test files and full execution logs are not passed directly to the coder. The coder then can revise the candidate and send it for another round of review.
Only the tester can approve final submission to prevent further coder modifications.
Appendix~\ref{app:harness_details} summarizes the operational limits for this extension. Both role prompts and controller messages are provided in the study repository.

\subsection{Self-verification failure analysis pipeline}
\label{sec:self_verification_pipeline}
We observed cases in which agents obtained passing results from their own checks but ultimately submitted implementations that failed the independent benchmark evaluation. This motivated an analysis of failed trajectories to identify deficiencies in how agents tested their implementations and interpreted or acted on the resulting evidence. Here, self-verification refers to the checks an agent performs to assess its implementation before submission. The analysis includes all 249 completed trials whose submitted implementations fail the benchmark evaluation: 146 from SimpleAgent and 103 from OpenHands. Trials that terminate without an evaluated submission are excluded.

\begin{figure}[h]
\begin{center}
\includegraphics[width=\linewidth, page=2]{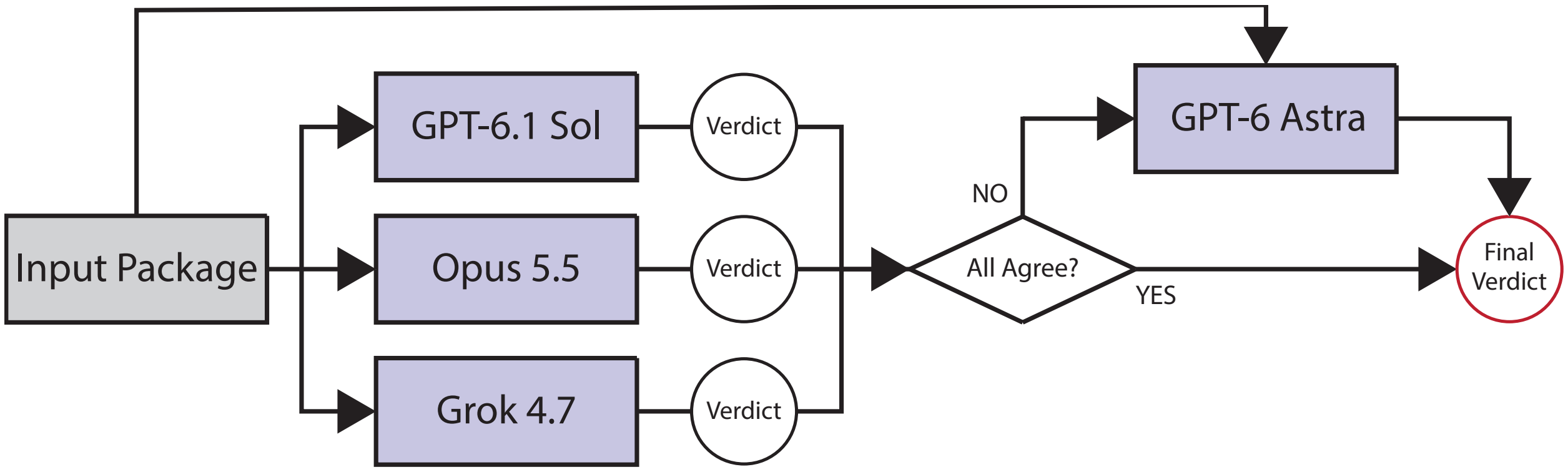}
\end{center}
\caption{\textbf{Overview of the LLM-as-judge workflow for categorizing self-verification failures.}
Three initial judges independently assess each input package using the codebook, assigning all applicable labels with a short rationale for each. Their label sets are then compared case by case. When all three judges assign identical label sets, the shared classification is retained as final. Cases with any label disagreement are passed to a final adjudicator. 
The final adjudicator uses the original input package, all three judges' verdicts and rationales, and the codebook to determine the final classification.}
\label{fig:judge_workflow}
\end{figure}

Codebook development combines manual review of more than 20 failed trajectories with pilot judging and iterative refinement. Targeted review of disagreements between two pilot judges helps clarify category boundaries and application criteria. We then apply the three-judge classification and adjudication procedure in Figure~\ref{fig:judge_workflow} to the full failure package.

\subsubsection{Codebook}
We develop the codebook iteratively from the inspected trajectories. Pilot LLM judgments help identify ambiguous category boundaries and behaviors that the current definitions do not adequately capture. An \texttt{OTHER} category allows judges to flag an identifiable verification failure outside the existing categories and explain the uncaptured behavior. We review these cases alongside the original trajectories and refine the category definitions, application criteria, and examples. We use Fleiss' $\kappa$~\citep{fleiss1971agreement} as an agreement criterion for deciding when to end codebook refinement. Appendix~\ref{app:judge_agreement} defines the statistic, and Table~\ref{tab:judge_agreement} reports agreement and assignment prevalence for the six verification-failure categories before adjudication.

The codebook records each category's identifier, name, description, application criteria, exclusion criteria, and illustrative examples. It defines six verification-failure categories: inadequate test configuration and coverage, generic consistency checks, ineffective test assertions, non-independent verification, test accommodation, and unverified post-test modification. It also includes \texttt{NV} for trajectories with no observable verification and \texttt{OTHER} for failures not adequately captured by the defined categories. The full codebook and judging prompts are provided in the study repository (\url{https://github.com/erfanhamdi/FEM-Bench-v2}).
A trajectory can receive multiple codes when each captures a distinct, supported deficiency. The \texttt{NV} label cannot co-occur with the six verification-failure categories. Judges are instructed to consider subsequent corrections and to avoid labeling an intermediate deficiency that was resolved before submission.

\subsubsection{LLM-as-judge}
For each trial, we construct an input package containing the original task prompt, the recorded trajectory, the submitted implementation, the reference implementation, and benchmark evaluation results. Trajectories from both harnesses are normalized into a sequence of recorded reasoning or messages, actions, and observations, including tool calls and execution feedback. Evaluation information includes the verification inputs, errors, and whether results match the reference. We omit raw evaluator output values and model metadata during preparation.

Three judges were used for the initial judgement steps, GPT-6.1 Sol, Claude Opus 5.5, and Grok 4.7. These initial judges are tasked to independently assess each package using the same codebook and judging instructions. Each returns all applicable category labels, a short evidence-based rationale for each label, and an uncertainty flag. The instructions require judges to examine the checks actually performed, their coverage and acceptance criteria, and how the agent uses the resulting evidence to revise or accept its implementation. Judges must distinguish planned tests from executed checks and agent-generated verification from independent benchmark evaluation. The configurations of the three judges and adjudicator, including output limits and retry settings, are documented in Appendix~\ref{app:judge_configuration} and Table~\ref{tab:judge_configuration}.

We compare the complete label sets assigned by the three judges. When all three sets are identical, we retain the shared classification. Any label disagreement sends the case to GPT-6 Astra for adjudication. Of the 249 cases, 104 have unanimous label sets and 145 undergo adjudication.

The adjudicator receives the original input package, the codebook, and all three judges' labels, rationales, and uncertainty flags. Coding model and judge identifiers are replaced with anonymous identifiers in the adjudication package. The adjudicator independently reassesses the evidence and may retain, remove, or add labels, including labels omitted by all initial judges. Its instructions explicitly prohibit majority voting or deference to a particular judge. We retain its classification as the final result for each adjudicated case.

\section{Results and Discussion}
\label{sec:results_discussion}

This section examines the performance and reliability of LLMs on FEM-Bench. We first evaluate single-pass implementation and test-generation performance (Section~\ref{sec:single_pass_results}). We then examine how execution and revision using an agentic harness affect implementation success, consistency across attempts, and cost (Section~\ref{sec:agentic_results}). We also present exploratory results from separating coding and testing into two agent roles (Section~\ref{sec:coder_tester_results}). Finally, we analyze self-verification failures in completed but incorrect agentic submissions and discuss their implications for scientific coding (Section~\ref{sec:self_verification_results}), before outlining the scope and limitations of our findings (Section~\ref{sec:scope_limitations}).

\subsection{Single-pass performance}
\label{sec:single_pass_results}
\begin{figure}[h]
\begin{center}
\includegraphics[width=\linewidth, page=2]{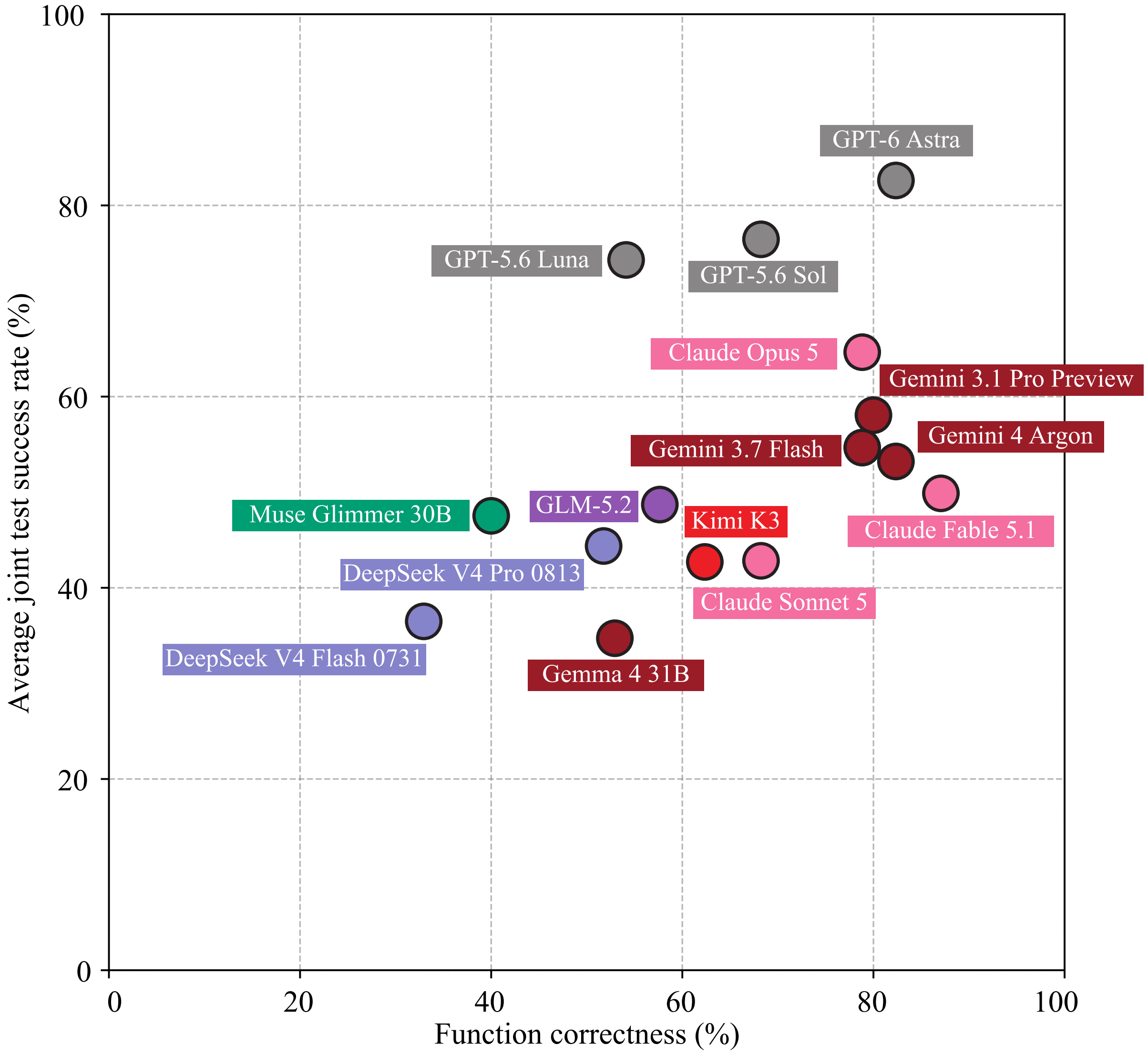}
\end{center}
\caption{\textbf{Single-pass performance on implementation and test-generation tasks.}
Each point represents one of the 15 evaluated models listed in Table~\ref{tab:models}. Results are computed separately for the implementation and test-generation tracks. Each track uses five trials per task (5 $\times$ 17 = 85 runs per model). The horizontal axis shows function correctness: the percentage of generated implementations that pass the benchmark evaluation. The vertical axis shows the average joint test success rate: the fraction of generated tests that both pass the reference implementation and reject the designated buggy implementations. Both of these metrics are averaged across tasks and trials. Higher values indicate better performance on each metric. The differing model rankings across the two axes show that stronger implementation performance does not necessarily mean stronger test-generation performance.}
\label{fig:single_pass_performance}
\end{figure}

We first evaluate the models' ability to implement finite element codes and generate tests in the single-pass setting. As described in Section~\ref{sec:single_pass_setup}, each trial consists of one generated response, without code execution or feedback that would allow the model to revise its output. The 15 models in Table~\ref{tab:models} each complete five trials on the 17 tasks in Table~\ref{tab:fem_bench_tasks}, giving 85 runs per model in each of the two tracks. For the implementation track, function correctness is the fraction of successful implementations averaged over tasks, equivalent to empirical Pass@1 (Equation~\ref{eq:function_correctness}). For the test-generation track, a test succeeds only if it accepts the reference implementation and rejects all expected failure cases (Equation~\ref{eq:joint_test_indicator}). We average the fraction of successful test cases within each trial and then across trials and tasks to obtain the joint test success rate (Equation~\ref{eq:joint_test_success}). Figure~\ref{fig:single_pass_performance} compares these two metrics. Appendix~\ref{app:additional_results} provides the model-level scores in Table~\ref{tab:single_pass_summary} and task-level implementation and test results in Table~\ref{tab:task_model_implementation_tests} and \ref{tab:task_model_test_generation_jts}.

Single-pass function correctness spans a wide range, from 32.9\% for DeepSeek V4 Flash 0731 to 87.1\% for Claude Fable 5.1. GPT-6 Astra and Gemini 4 Argon both achieve 82.4\%, followed by Gemini 3.1 Pro Preview at 80.0\%. These results demonstrate that several models can produce successful implementations for a substantial fraction of attempts without execution feedback. Performance nevertheless varies considerably across tasks (Table~\ref{tab:task_model_implementation_tests}). For example, 14 of the 15 models solve the Delaunay circular-mesh task in all five trials. In contrast, the Specht shape-function task is solved in only four of the 75 trials across models, with two successes each from Claude Fable 5.1 and Gemini 4 Argon. Every task is solved at least once by at least one model, but no model succeeds on every task in all five trials.

The test-generation track produces a different model ranking. Joint test success ranges from 34.7\% for Gemma 4 31B to 82.6\% for GPT-6 Astra, with GPT-5.6 Sol and GPT-5.6 Luna reaching 76.5\% and 74.3\%, respectively (Table~\ref{tab:single_pass_summary}). Claude Fable 5.1, which leads in function correctness, achieves 49.9\% joint test success. Conversely, GPT-5.6 Luna achieves 54.1\% function correctness while ranking third in test generation. Thus, stronger implementation performance does not necessarily imply stronger performance in generating tests that distinguish correct from incorrect code. Because the two tracks use different success criteria and aggregation rules, their percentages should be interpreted as complementary measures rather than interchangeable scores.

\subsection{Agentic performance}
\label{sec:agentic_results}

\begin{figure}[h]
	\begin{center}
		\includegraphics[width=\linewidth, page=2]{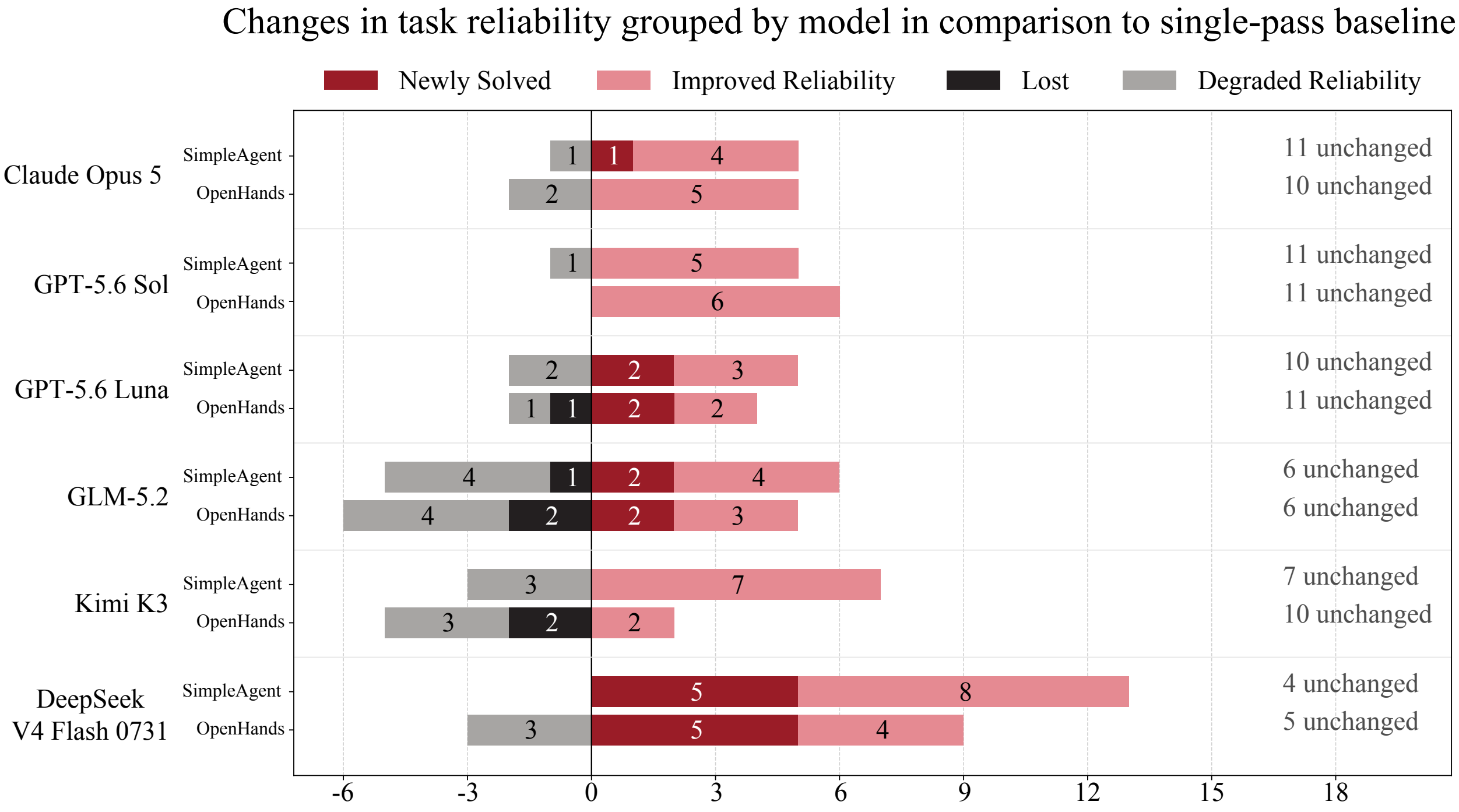}
	\end{center}
	\caption{\textbf{Changes in task reliability relative to the single-pass baseline.}
	Each row represents one model--harness combination across 17 tasks, with reliability measured by the number of successful runs out of five trials. Tasks are classified as newly solved when the success count increases from zero to a positive value, and as improved reliability when it increases from an already positive value. Degraded reliability indicates a decrease that remains above zero, whereas lost tasks have a positive baseline success count that falls to zero. Improvements extend to the right and regressions to the left. Unchanged task counts are shown separately. Most gains reflect improved reliability on previously solved tasks.}
\label{fig:agentic_reliability_changes}
\end{figure}

\begin{figure}[h]
\begin{center}
\includegraphics[width=\linewidth, page=3]{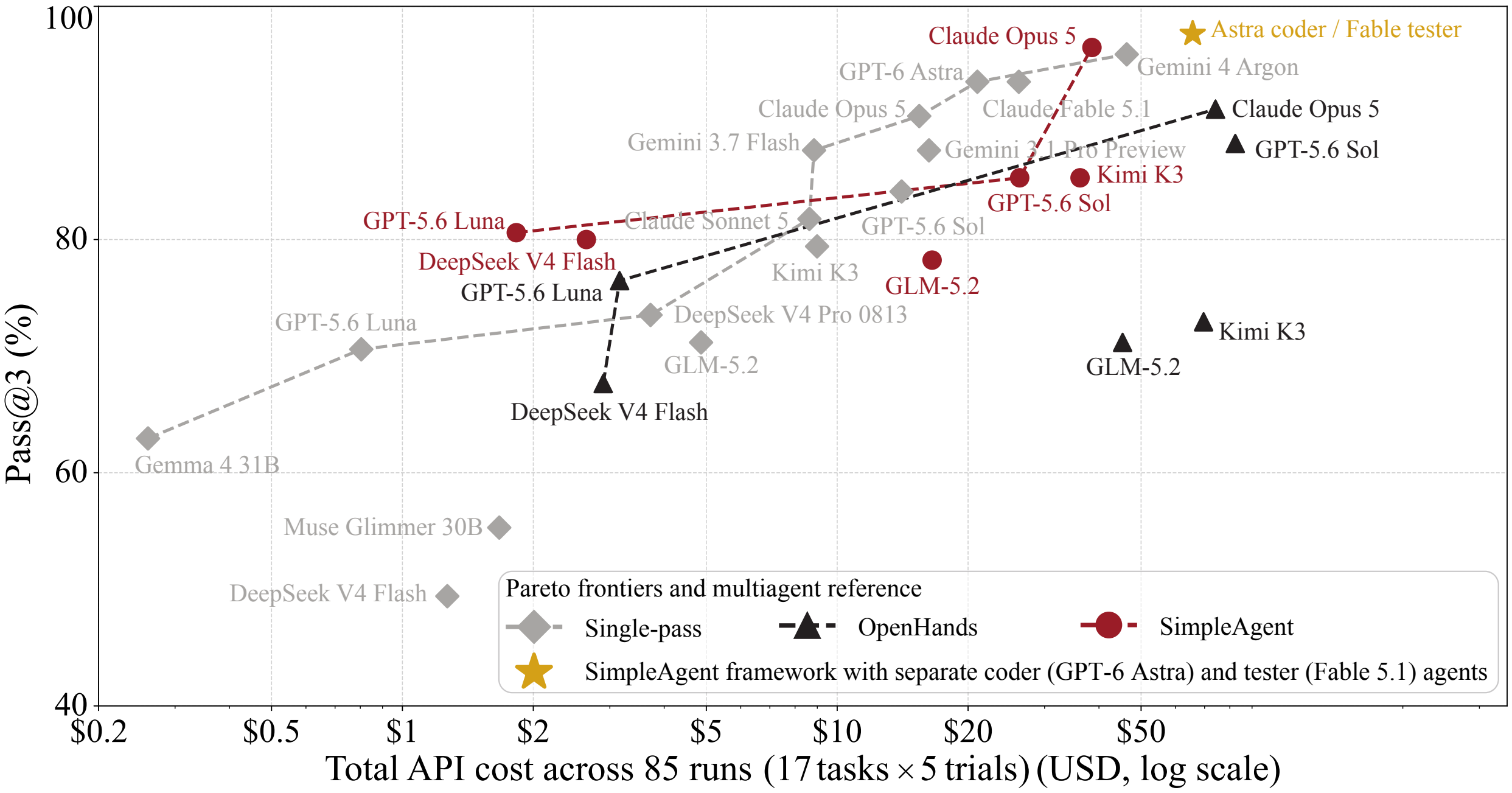}
\end{center}
\caption{\textbf{Pass@3 performance versus recorded cost.}
Pass@3 is estimated from five trials per task and averaged across 17 tasks. The horizontal axis shows the total recorded cost across all 85 runs on a logarithmic scale. Dashed lines connect the Pareto-optimal configurations within the single-pass, SimpleAgent, and OpenHands settings. The star denotes a multi-agent configuration built on SimpleAgent, with GPT-6 Astra as the coder and Claude Fable 5.1 as the tester. SimpleAgent offers competitive performance at lower recorded cost than OpenHands for the evaluated models, while the coder-tester configuration achieves the highest Pass@3 among the configurations shown.}
\label{fig:agentic_pass_at_3_cost}
\end{figure}

\begin{figure}[h]
\begin{center}
\includegraphics[width=\linewidth, page=2]{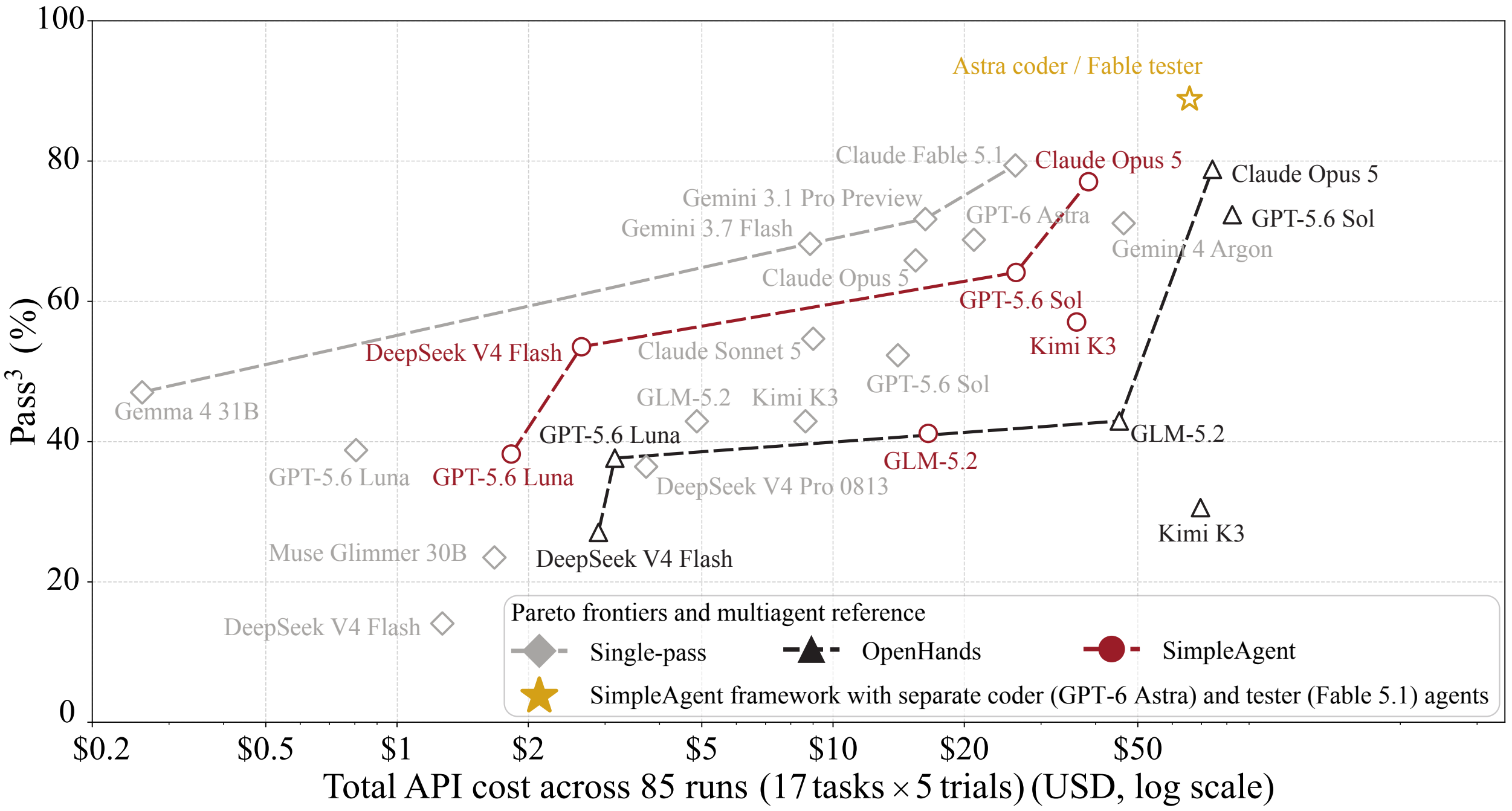}
\end{center}
\caption{\textbf{Pass$^3$ performance versus recorded cost.}
Pass$^3$ estimates the probability that all three attempts succeed, using five trials per task and averaging across 17 tasks. The horizontal axis shows the total recorded cost across all 85 runs on a logarithmic scale. Dashed lines connect the Pareto-optimal configurations within the single-pass, SimpleAgent, and OpenHands settings. The star denotes a multi-agent configuration built on SimpleAgent, with GPT-6 Astra as the coder and Claude Fable 5.1 as the tester. SimpleAgent achieves higher Pass$^3$ than OpenHands for three of the six shared models at lower recorded cost, while the coder-tester configuration achieves the highest Pass$^3$ among the configurations shown.}
\label{fig:agentic_pass_power_3_cost}
\end{figure}

We compare SimpleAgent and OpenHands with the single-pass baseline for six models: Claude Opus 5, GPT-5.6 Sol, GPT-5.6 Luna, Kimi K3, GLM-5.2, and DeepSeek V4 Flash 0731 (Table~\ref{tab:models}). Each model is evaluated on the implementation track using the same 17 tasks and five trials per task in each setting. Figure~\ref{fig:agentic_reliability_changes} shows changes in task-level reliability, while Figures~\ref{fig:agentic_pass_at_3_cost} and~\ref{fig:agentic_pass_power_3_cost} compare performance and recorded cost with the single-pass results. Here, reliability describes consistency across repeated trials, measured for each task by its number of successful implementations out of five. A task is newly solved if this count increases from zero to a positive value, and its reliability improves if the count increases from an already positive value. A decrease is classified as degraded reliability if at least one success remains, or as a lost task if the count falls to zero. Tables~\ref{tab:task_model_implementation_tests} and~\ref{tab:reliability_changes_summary} provide the task-level outcomes and category counts.

Most gains come from improved reliability on tasks that already have successful single-pass trials. Across the 102 model-task combinations for each harness, SimpleAgent improves reliability on 31 and newly solves 10, while OpenHands improves reliability on 22 and newly solves nine (Figure~\ref{fig:agentic_reliability_changes}). DeepSeek V4 Flash 0731 shows the largest increase in function correctness with SimpleAgent, from 32.9\% to 65.9\%, compared with 44.7\% under OpenHands. With SimpleAgent, it newly solves five tasks and improves reliability on eight, with no regressions. Its one-dimensional linear elastic dynamics task, for example, changes from zero successful single-pass trials to five successful SimpleAgent trials. Claude Opus 5 also newly solves the Specht shape-function task with SimpleAgent, succeeding in two of five trials after no single-pass successes. Thus, execution and revision can enable a model to solve tasks on which none of its observed single-pass attempts succeeded.

These improvements are accompanied by regressions. SimpleAgent loses one model-task combination and degrades reliability on 11, compared with five lost and 13 degraded combinations for OpenHands. Some losses involve budget exhaustion: Kimi K3 succeeds in all five single-pass trials on the cohesive-zone T2 task, but all five OpenHands trials exhaust the token budget without an evaluated submission. Across the full agentic evaluation, nine of 510 SimpleAgent trials and 83 of 510 OpenHands trials terminate at a token or wall-time limit. Two additional OpenHands trials ended without producing a submission and were scored as failures. However, budget exhaustion does not explain every loss. GPT-5.6 Luna falls from five single-pass successes to zero with OpenHands on three-dimensional linear elastic dynamics with supplied helpers, despite completing all five runs and submitting implementations for evaluation. These outcomes distinguish failure to produce an evaluated submission within the budget from failure of a completed implementation. Both types of failure contribute zero successes to the reported metrics.

Figures~\ref{fig:agentic_pass_at_3_cost} and~\ref{fig:agentic_pass_power_3_cost} summarize the results using Pass@3 and Pass$^3$, estimated from the five trials and averaged equally over tasks according to Equations~\ref{eq:pass_at_3} and~\ref{eq:pass_power_3}. SimpleAgent increases Pass@3 over the single-pass baseline for all six models and exceeds OpenHands for five of them (Table~\ref{tab:implementation_pass_at_3_by_method}). Its advantage is less uniform for Pass$^3$. SimpleAgent improves over single-pass generation for four models and exceeds OpenHands for three (Table~\ref{tab:implementation_pass_power_3_by_method}). GPT-5.6 Sol performs better with OpenHands on both metrics, reaching 88.2\% Pass@3 and 72.4\% Pass$^3$, compared with 85.3\% and 64.1\% with SimpleAgent. OpenHands also gives Claude Opus 5 a slightly higher Pass$^3$, at 78.8\% versus 77.1\%. The relative performance of the harnesses therefore depends on both the model and the metric.

The distinction between Pass@3 and Pass$^3$ is important when interpreting scores near 100\%. Pass@3 measures the chance of obtaining at least one successful implementation in three complete trials, whereas Pass$^3$ requires all three to succeed. Pass@3 is relevant when several candidates can be generated and independently checked. Pass$^3$ captures consistency when repeatedly using the same system. Claude Opus 5 with SimpleAgent achieves 96.5\% Pass@3 but 77.1\% Pass$^3$. Across the six models and two harnesses, the highest Pass$^3$ is 78.8\%, achieved by Claude Opus 5 with OpenHands. Pass$^3$ cannot exceed Pass@3 by definition. The observed gap shows that a high chance of success within three attempts can still coexist with failures across repeated trials. For GPT-5.6 Luna, SimpleAgent increases Pass@3 from 70.6\% to 80.6\%. Its Pass$^3$ changes from 38.8\% to 38.2\%. Here, the higher chance of occasional success does not come with improved consistency.

The Pareto frontiers in Figures~\ref{fig:agentic_pass_at_3_cost} and~\ref{fig:agentic_pass_power_3_cost} identify the best observed tradeoffs between performance and recorded cost within each setting. Cost is summed over the 85 runs for each configuration and shown on a logarithmic axis. Reported costs account for prompt caching and include estimates where usage records are incomplete. SimpleAgent has lower recorded cost than OpenHands for all six shared models. For Claude Opus 5, the totals are \$38.54 and \$74.15, respectively, with SimpleAgent achieving higher Pass@3 and slightly lower Pass$^3$. Single-pass configurations remain competitive at high Pass$^3$.

\subsubsection{Exploratory coder-tester results}
\label{sec:coder_tester_results}
The exploratory configuration with GPT-6 Astra as the coder and Claude Fable 5.1 as the tester is shown by the stars in Figures~\ref{fig:agentic_pass_at_3_cost} and~\ref{fig:agentic_pass_power_3_cost}. It achieves the highest observed Pass@3 and Pass$^3$ among the evaluated configurations: 97.6\% and 88.8\%, respectively. Its total recorded cost is \$65.72. It succeeds in all five trials on 15 of the 17 tasks (Table~\ref{tab:task_model_implementation_tests}). Performance also remains high when the roles are reversed, with Fable as coder and Astra as tester achieving 91.8\% Pass@3 and 84.7\% Pass$^3$. All 170 coder–tester trials produced submissions. Six Astra/Fable and ten Fable/Astra submissions failed independent evaluation. None of the trials terminated because of resource limits.

The coder-tester design introduces a verification process external to the coder's own implementation and self-testing. As described in Section~\ref{sec:coder_tester_setup}, the tester develops tests from the task specification without access to the coder's conversation history or implementation source. It therefore serves as an approximate external oracle, constructing tests with expected outcomes and supplying diagnostic feedback when a candidate fails them. This gives the coder new evidence with which to revisit assumptions and revise its implementation. Retaining the test suite allows the tester to check that successive candidates resolve earlier failures. Assigning final approval to the tester requires the candidate to satisfy these checks before submission.

A hyperelasticity trajectory illustrates how this feedback can support successive corrections. For the T3 2D hyperelasticity task, Fable's initial candidate reproduced an analytical homogeneous deformation with a maximum displacement discrepancy of approximately $1.4\times10^{-11}$. Astra nevertheless found failures under additional loading and mesh configurations. Its feedback first prompted a correction to the tangent calculation, reducing failures in the existing 15-test suite from seven to four. Further tests exposed a violation of rigid-translation invariance: imposing a uniform boundary translation produced an incorrect displacement at the interior node. Fable traced this failure to an incorrect shape-function derivative and corrected it in the third candidate. The final implementation passed all 27 accumulated tester checks, with no tests retired, and passed the independent FEM-Bench evaluation. This trajectory shows how broader testing and successive feedback can expose and correct numerical errors that survive the coder's initial validation.

The Pass$^3$ results make the reliability implications particularly clear. Astra as coder with Fable as tester reaches 88.8\%, compared with single-pass scores of 68.8\% for Astra and 79.4\% for Fable (Table~\ref{tab:implementation_pass_power_3_by_method}). The reversed pairing also exceeds both single-pass scores, reaching 84.7\%. Because Pass$^3$ estimates the probability that all three attempts succeed (Equation~\ref{eq:pass_power_3}), these results indicate greater consistency of correct implementations across repeated trials under the benchmark criterion. Together with the trajectory above, this highlights the value of external verification and actionable feedback for reliable scientific coding: a separate tester can expose numerical errors missed by the coder's own checks and guide corrections before final submission.

\subsection{Self-verification failure analysis}
\label{sec:self_verification_results}
\begin{figure}[h]
\begin{center}
\includegraphics[width=\linewidth, page=2]{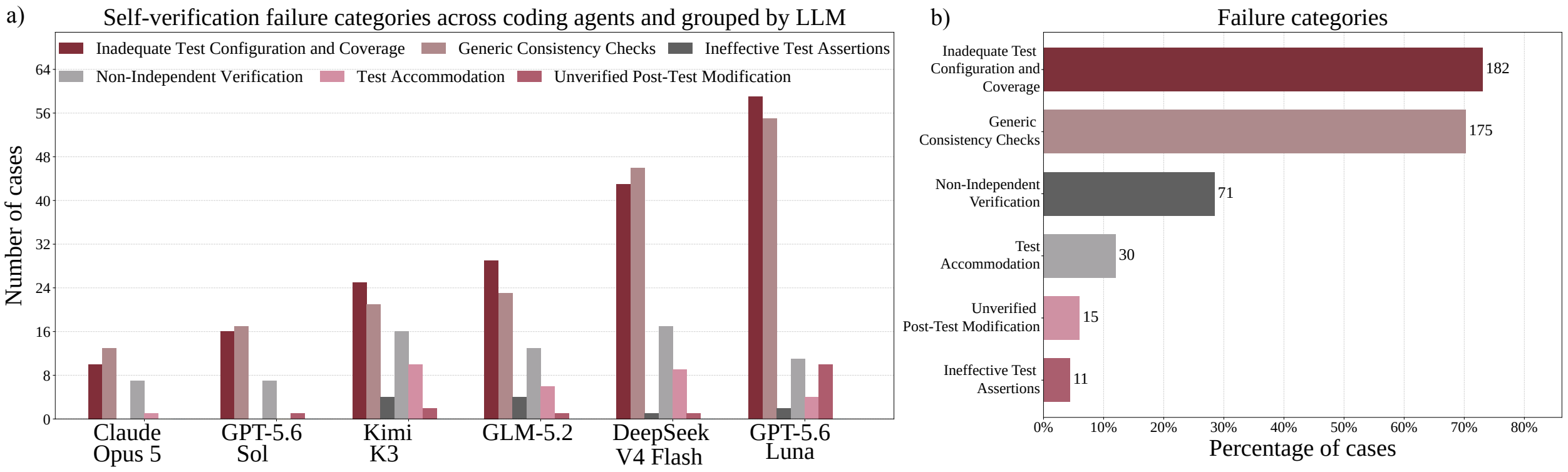}
\end{center}
\caption{\textbf{Self-verification failure categories among completed failed submissions.} (a) Categories grouped by model.
Each bar shows the number of analyzed failed cases assigned to a particular category, pooled across SimpleAgent and OpenHands. Models are ordered by increasing number of analyzed failed cases. Inadequate Test Configuration and Coverage and Generic Consistency Checks are the most frequently assigned categories. A case may receive multiple labels, so category counts can sum to more than the number of cases for a model.
(b) Distribution of self-verification failure categories.
 Bars show the percentage of the 249 analyzed failed cases assigned to each category, pooled across models and agentic harnesses. Categories are ordered by decreasing frequency, and numbers beside the bars indicate case counts. A case may receive multiple labels, so the percentages do not sum to 100\%.}
\label{fig:self_verification_failures}
\end{figure}

We analyze the verification behavior in 249 completed agentic trials whose submitted implementations fail the independent benchmark evaluation: 146 from SimpleAgent and 103 from OpenHands. The procedure in Section~\ref{sec:self_verification_pipeline} identifies weaknesses in the checks performed, the evidence used to accept a candidate, and the handling of revisions before submission. Figure~\ref{fig:self_verification_failures}(a) and Table~\ref{tab:failure_categories_by_model} show the category counts by model. Figure~\ref{fig:self_verification_failures}(b) and Table~\ref{tab:failure_category_distribution} show their overall distribution and counts by harness. Labels can overlap. The percentages refer only to these 249 failed final submissions. Successful submissions and runs that ended without a submission, including those stopped by budget exhaustion, are excluded. The 94 excluded trials comprise 91 token-limit terminations, one wall-time termination, and two missing submissions.

The dominant categories are Inadequate Test Configuration and Coverage (VF-01), assigned to 182 cases (73.1\%), and Generic Consistency Checks (VF-02), assigned to 175 (70.3\%). They co-occur in 137 cases (55.0\%) and are observed for all six models and both harnesses. These categories often occur together: an agent tests only a limited set of cases, and its checks are too general to verify the numerical results. This highlights both the choice of test cases and the criteria used to judge their results.

Iterative refinement of the codebook yielded six categories of self-verification failure. We describe each category below and illustrate it with an example from the recorded trajectories.

\paragraph{VF-01: Inadequate Test Configuration and Coverage.}
This code identifies verification that omits relevant geometries, loading regimes, outputs, or state variables. For example, GPT-5.6 Sol tested its implementation of the 3D linear elastic dynamics task (T3) exclusively on an axis-aligned unit cube. Its tests passed, but the submitted solver failed the distorted-mesh case in the evaluator.

\paragraph{VF-02: Generic Consistency Checks.}
This code captures reliance on general properties as sufficient evidence of numerical correctness. For example, in the 3D linear elastic dynamics task (T3), GLM-5.2 treated stiffness-matrix symmetry, rigid-body responses, and the unit-cube Jacobian determinant as confirmation that the formulation was correct. The submitted solver nevertheless failed the distorted-mesh case in the evaluator.

VF-01 concerns the aspects of the solver that are left untested, and VF-02 concerns checks that are too general to establish correctness. Both can occur in the same trial. General consistency checks can expose particular errors. Passing them does not establish correctness of the full numerical formulation.

\paragraph{VF-03: Ineffective Test Assertions.}
In 11 cases (4.4\%), assertions were too permissive to detect meaningful numerical errors. For example, in the 3D linear elastic dynamics task (T3), GPT-5.6 Luna used an absolute tolerance larger than the expected displacement, so even a zero response would pass. Its tests passed, but the submitted solver failed all six benchmark checks.

\paragraph{VF-04: Non-Independent Verification.}
In 71 cases (28.5\%), the implementation and its reference calculations shared an unvalidated assumption or error. For example, in the 1D isotropic-hardening task, Kimi K3 used the same incorrect interpretation of a material parameter in its implementation and expected test results. Its tests passed, but four of five benchmark checks failed.

\paragraph{VF-05: Test Accommodation.}
In 30 cases (12.0\%), the agent changed the test or its interpretation of the result to accommodate a discrepancy. It did not correct the implementation. For example, in the 1D kinematic-hardening task, Kimi K3 replaced a correct expected stress with a calculation that matched its incorrect output. The revised tests passed, but the unchanged implementation failed three of five benchmark checks.

\paragraph{VF-06: Unverified Post-Test Modification.}
In 15 cases (6.0\%), the agent modified an implementation after testing and submitted it without checking the revised version. For example, GPT-5.6 Luna changed the plastic-strain update after passing tests for the 1D kinematic-hardening task. The untested revision produced negative accumulated plastic strain in the evaluator.

These failures show that a simulation can run successfully and produce plausible results while still being incorrect. Passing the agent’s own tests is therefore not enough to confirm that the computation is correct.

These findings motivate a verification component external to the coder's own implementation process. Its value depends on testing task-relevant conditions and supplying independently justified expected results. It must also use meaningful numerical tolerances and preserve valid tests when they expose a discrepancy. The multi-agent trajectory discussed in Section~\ref{sec:coder_tester_results} illustrates this corrective role. In that trajectory the tester identified failures missed by the coder, retained its checks, and supplied feedback that led to successive numerical corrections. The Astra/Fable configuration's 88.8\% Pass$^3$ (Figure~\ref{fig:agentic_pass_power_3_cost}) provides complementary evidence of strong repeated-trial reliability. Together, the failure categories and this example highlight why autonomous scientific agents need an effective external oracle for testing and acceptance, with expectations grounded in the task's scientific requirements.

The code an agent submits should be the same version that passed its tests. Fourteen of the 15 VF-06 cases occur in SimpleAgent (Table~\ref{tab:failure_category_distribution}). Its submission gate checks the latest execution's exit code. It does not ensure that the submitted code is the version that was tested. The coder-tester extension addresses this specific gap by assigning the submission role to the tester agent. The tester is tasked with running the retained test suite against a candidate and deciding whether to submit it. Scientific reliability therefore depends both on the quality of the oracle and on preserving the connection between the evidence obtained and the code ultimately used.

\subsection{Scope and limitations}
\label{sec:scope_limitations}

This study is limited to 17 finite element tasks, with five trials per task (Table~\ref{tab:fem_bench_tasks}). Agentic evaluations use fixed token budgets (Table~\ref{tab:experimental_configurations}). The cost of repeated execution and revision also limits the number of models and configurations that can be tested.
Within these experiments, the harnesses differ in their prompts, tools, and submission rules. Their comparison therefore does not isolate individual design choices. 
The test-generation track evaluates separately prompted tests on predefined reference and faulty implementations (Section~\ref{sec:evaluation_procedure}). Its scores are limited to those cases and do not directly measure agents' ability to detect errors in their own code. We examined agents' own checks through the analysis of failed agentic trajectories. Manual review informed the codebook used in this analysis, but LLM judging and adjudication may still misclassify cases (Section~\ref{sec:self_verification_pipeline}). Other tasks or workflows may reveal additional failure categories.

\section{Conclusion}
\label{sec:conclusion}
We extended FEM-Bench to 17 demanding finite element tasks and evaluated LLMs in single-pass and agentic settings, alongside separate test-generation and self-verification analyses. Several models demonstrated strong implementation performance without execution feedback, with the best single-pass model succeeding in 87.1\% of trials. Allowing execution and revision often improved consistency and enabled success on tasks with no successful single-pass trials. These gains varied across models and harnesses, and regressions also occurred. The results demonstrate substantial capability for implementing scientific computations while identifying remaining challenges in consistent correctness and verification.

Reliable scientific coding requires both consistent implementation success and checks that can detect numerical errors. A high probability of obtaining a correct implementation within several attempts does not establish consistent success or the ability to identify the correct candidate. Model rankings differed between implementation and test generation, showing that stronger performance in one did not necessarily imply stronger performance in the other. Our analysis of 249 incorrect agentic submissions identified checks with inadequate coverage, generic acceptance criteria, or incorrect assumptions shared with the implementation. These observations show why passing an agent's own tests does not provide sufficient evidence of numerical correctness.

The exploratory coder-tester configuration achieved the highest observed Pass@3 and Pass$^3$, and a recorded trajectory illustrated how a separate tester could expose numerical errors and guide successive corrections. Our findings highlight the importance of an external oracle that provides independently justified numerical checks and diagnostic feedback to help agents detect and correct implementation errors.

Future work should extend FEM-Bench with checks based on known solutions, convergence studies, and physical principles. Agents' tests could be evaluated by whether they accept valid implementations and detect deliberately introduced errors. Comparing agents with and without reusable verification tools could help distinguish difficulties in building checks from difficulties in choosing them. Separate coder and verifier agents should also be compared with single-agent systems using the same compute budget, measuring correctness, consistency, and cost. As LLMs take on greater responsibility for scientific computation, their usefulness will increasingly depend on providing trustworthy evidence that their numerical results are correct.

\subsubsection*{Acknowledgments}
The authors gratefully acknowledge support from the Boston University Department of Mechanical Engineering, the
National Science Foundation through the U.S. National Science Foundation Cyberinfrastructure for Sustained Scientific Innovation program, under the “Elements: Curating and Disseminating Solid Mechanics Based Benchmark Datasets” (Project CMMI-2310771), the U.S. Department of Energy, Advanced Scientific Computing Research program, under the “Resolution-invariant deep learning for accelerated propagation of epistemic and aleatory uncertainty in multi-scale energy storage systems, and beyond” project (Project No. 81824), the Amazon Research
Award program, and the BU SAIL AI Pipeline Pilot.
\bibliographystyle{unsrtnat}
\bibliography{references}

\clearpage
\appendix
\section{Additional results}
\label{app:additional_results}
\begin{table}[!htbp]
\centering
\caption{\textbf{Single-pass implementation and test-generation performance.} Function implementation measures the fraction of submissions that pass the benchmark evaluation. Joint test success rate measures the fraction of generated tests that accept the reference implementation and reject all designated faulty implementations. Both metrics follow Section~\ref{sec:evaluation_metrics} and are percentages averaged over five trials per task, with equal weight across 17 tasks. Bold marks the highest value in each column.}
\label{tab:single_pass_summary}
\small
\renewcommand{\arraystretch}{1.1}
\setlength{\tabcolsep}{4pt}

\begin{tabular}{@{}l c c@{}}
\toprule
\textbf{Model} & \textbf{\shortstack{Function\\implementation (\%)}} & \textbf{\shortstack{Joint test\\success rate (\%)}} \\
\midrule
Claude Fable 5.1 & \textbf{87.1} & 49.9 \\
Claude Opus 5 & 78.8 & 64.6 \\
Claude Sonnet 5 & 68.2 & 42.8 \\
GPT-6 Astra & 82.4 & \textbf{82.6} \\
GPT-5.6 Sol & 68.2 & 76.5 \\
GPT-5.6 Luna & 54.1 & 74.3 \\
Gemini 3.1 Pro Preview & 80.0 & 58.1 \\
Gemini 3.7 Flash & 78.8 & 54.7 \\
Gemini 4 Argon & 82.4 & 53.2 \\
Gemma 4 31B & 52.9 & 34.7 \\
Kimi K3 & 62.4 & 42.7 \\
GLM-5.2 & 57.6 & 48.7 \\
DeepSeek V4 Flash 0731 & 32.9 & 36.5 \\
DeepSeek V4 Pro 0813 & 51.8 & 44.4 \\
Muse Glimmer 30B & 40.0 & 47.5 \\
\bottomrule
\end{tabular}
\end{table}

\begin{table}[!htbp]
\centering
\caption{\textbf{Changes in task reliability relative to single-pass generation.} Each task has five implementation trials. Newly solved tasks increase from zero to at least one successful trial. Improved tasks increase from a positive success count. Lost tasks fall to zero successes, while degraded tasks decrease but retain at least one success. Unchanged tasks have the same success count. Entries count tasks, and each row sums to 17.}
\label{tab:reliability_changes_summary}
\small
\renewcommand{\arraystretch}{1.1}
\setlength{\tabcolsep}{4pt}

\begin{tabularx}{\linewidth}{@{}l l *{5}{>{\centering\arraybackslash}X}@{}}
\toprule
\textbf{Model} & \textbf{Method} & \textbf{\mbox{Newly solved}} & \textbf{Improved} & \textbf{Lost} & \textbf{Degraded} & \textbf{Unchanged} \\
\midrule
\multirow{2}{*}{Claude Opus 5} & SimpleAgent & 1 & 4 & 0 & 1 & 11 \\
 & OpenHands & 0 & 5 & 0 & 2 & 10 \\
\arrayrulecolor{gray!70}\midrule\arrayrulecolor{black}
\multirow{2}{*}{GPT-5.6 Sol} & SimpleAgent & 0 & 5 & 0 & 1 & 11 \\
 & OpenHands & 0 & 6 & 0 & 0 & 11 \\
\arrayrulecolor{gray!70}\midrule\arrayrulecolor{black}
\multirow{2}{*}{GPT-5.6 Luna} & SimpleAgent & 2 & 3 & 0 & 2 & 10 \\
 & OpenHands & 2 & 2 & 1 & 1 & 11 \\
\arrayrulecolor{gray!70}\midrule\arrayrulecolor{black}
\multirow{2}{*}{Kimi K3} & SimpleAgent & 0 & 7 & 0 & 3 & 7 \\
 & OpenHands & 0 & 2 & 2 & 3 & 10 \\
\arrayrulecolor{gray!70}\midrule\arrayrulecolor{black}
\multirow{2}{*}{GLM-5.2} & SimpleAgent & 2 & 4 & 1 & 4 & 6 \\
 & OpenHands & 2 & 3 & 2 & 4 & 6 \\
\arrayrulecolor{gray!70}\midrule\arrayrulecolor{black}
\multirow{2}{*}{DeepSeek V4 Flash 0731} & SimpleAgent & 5 & 8 & 0 & 0 & 4 \\
 & OpenHands & 5 & 4 & 0 & 3 & 5 \\
\bottomrule
\end{tabularx}
\end{table}
\begin{table}[!htbp]
\centering
\caption{\textbf{Pass@3 (implementation) by model and method.} Pass@3 estimates the probability of at least one successful implementation in three attempts (Equation~\ref{eq:pass_at_3}). Values are percentages estimated from five trials per task and averaged equally across 17 tasks. Dashes indicate unevaluated combinations. Bold marks the highest score in each method column.}
\label{tab:implementation_pass_at_3_by_method}
\small
\renewcommand{\arraystretch}{1.1}
\setlength{\tabcolsep}{4pt}

\begin{tabular}{@{}l c c c@{}}
\toprule
\textbf{Model} & \textbf{Single-pass} & \textbf{SimpleAgent} & \textbf{OpenHands} \\
\midrule
Claude Fable 5.1 & 93.5 & --- & --- \\
Claude Opus 5 & 90.6 & \textbf{96.5} & \textbf{91.2} \\
Claude Sonnet 5 & 79.4 & --- & --- \\
GPT-6 Astra & 93.5 & --- & --- \\
GPT-5.6 Sol & 84.1 & 85.3 & 88.2 \\
GPT-5.6 Luna & 70.6 & 80.6 & 76.5 \\
Gemini 3.1 Pro Preview & 87.6 & --- & --- \\
Gemini 3.7 Flash & 87.6 & --- & --- \\
Gemini 4 Argon & \textbf{95.9} & --- & --- \\
Gemma 4 31B & 62.9 & --- & --- \\
Kimi K3 & 81.8 & 85.3 & 72.9 \\
GLM-5.2 & 71.2 & 78.2 & 71.2 \\
DeepSeek V4 Flash 0731 & 49.4 & 80.0 & 67.6 \\
DeepSeek V4 Pro 0813 & 73.5 & --- & --- \\
Muse Glimmer 30B & 55.3 & --- & --- \\
\bottomrule
\end{tabular}
\end{table}

\begin{table}[!htbp]
\centering
\caption{\textbf{Pass$^3$ (implementation) by model and method.} Pass$^3$ estimates the probability that all three implementations succeed in three attempts (Equation~\ref{eq:pass_power_3}). Values are percentages estimated from five trials per task and averaged equally across 17 tasks. Dashes indicate unevaluated combinations. Bold marks the highest score in each method column.}
\label{tab:implementation_pass_power_3_by_method}
\small
\renewcommand{\arraystretch}{1.1}
\setlength{\tabcolsep}{4pt}

\begin{tabular}{@{}l c c c@{}}
\toprule
\textbf{Model} & \textbf{Single-pass} & \textbf{SimpleAgent} & \textbf{OpenHands} \\
\midrule
Claude Fable 5.1 & \textbf{79.4} & --- & --- \\
Claude Opus 5 & 65.9 & \textbf{77.1} & \textbf{78.8} \\
Claude Sonnet 5 & 54.7 & --- & --- \\
GPT-6 Astra & 68.8 & --- & --- \\
GPT-5.6 Sol & 52.4 & 64.1 & 72.4 \\
GPT-5.6 Luna & 38.8 & 38.2 & 37.6 \\
Gemini 3.1 Pro Preview & 71.8 & --- & --- \\
Gemini 3.7 Flash & 68.2 & --- & --- \\
Gemini 4 Argon & 71.2 & --- & --- \\
Gemma 4 31B & 47.1 & --- & --- \\
Kimi K3 & 42.9 & 57.1 & 30.6 \\
GLM-5.2 & 42.9 & 41.2 & 42.9 \\
DeepSeek V4 Flash 0731 & 14.1 & 53.5 & 27.1 \\
DeepSeek V4 Pro 0813 & 36.5 & --- & --- \\
Muse Glimmer 30B & 23.5 & --- & --- \\
\bottomrule
\end{tabular}
\end{table}

\begin{sidewaystable}[p]
\centering
\caption{\textbf{Implementation success out of five trials.} Each task entry counts implementations that pass the benchmark evaluation. Entries follow the method order in the headers. Summary metrics give equal weight to all 17 tasks. Task abbreviations follow Table~\ref{tab:fem_bench_tasks}.}
\label{tab:task_model_implementation_tests}
\fontsize{8}{10}\selectfont
\renewcommand{\arraystretch}{2.4}
\setlength{\tabcolsep}{1.5pt}
\setlength{\dashlinedash}{2pt}
\setlength{\dashlinegap}{2pt}
\ADLactivate

\resizebox{\linewidth}{!}{%
\begin{tabular}{@{}l*{9}{c}@{\hspace{6pt}}>{\arrayrulecolor{gray}}*{5}{>{\hspace{1.5pt}}c<{\hspace{1.5pt}}:}>{\hspace{1.5pt}}c<{\hspace{1.5pt}}@{\hspace{8pt}}>{\arrayrulecolor{black}}c@{\hspace{6pt}}c@{}}
\toprule
 & \multicolumn{9}{c}{\textbf{Single-pass (Implementation)}} & \multicolumn{6}{c}{\textbf{Single-pass, SimpleAgent, OpenHands (Implementation)}} & \multicolumn{2}{c}{\textbf{Multiagent}} \\
\cmidrule(lr){2-10}\cmidrule(lr){11-16}\cmidrule(lr){17-18}
\textbf{Task} & \shortstack{Claude\\Fable 5.1} & \shortstack{Claude\\Sonnet 5} & \shortstack{GPT-6\\Astra} & \shortstack{Gemini\\3.1 Pro\\Preview} & \shortstack{Gemini\\3.7 Flash} & \shortstack{Gemini\\4 Argon} & \shortstack{Gemma\\4 31B} & \shortstack{DeepSeek\\V4 Pro\\0813} & \shortstack{Muse\\Glimmer\\30B} & \shortstack{Claude\\Opus 5} & \shortstack{GPT-5.6\\Sol} & \shortstack{GPT-5.6\\Luna} & \shortstack{Kimi K3} & \shortstack{GLM-5.2} & \shortstack{DeepSeek\\V4 Flash\\0731} & \shortstack{Fable (coder)\\ Astra (tester)} & \shortstack{Astra (coder)\\Fable (tester)} \\
\midrule
1D isotropic hardening & 5/5 & 4/5 & 5/5 & 5/5 & 5/5 & 5/5 & 1/5 & 5/5 & 4/5 & 5/5, 5/5, 5/5 & 5/5, 5/5, 5/5 & 5/5, 5/5, 5/5 & 4/5, 4/5, 2/5 & 5/5, 4/5, 5/5 & 3/5, 5/5, 4/5 & 5/5 & 5/5 \\
1D kinematic hardening & 5/5 & 4/5 & 5/5 & 5/5 & 5/5 & 2/5 & 5/5 & 5/5 & 5/5 & 5/5, 5/5, 5/5 & 2/5, 1/5, 3/5 & 2/5, 2/5, 2/5 & 3/5, 4/5, 3/5 & 2/5, 2/5, 5/5 & 4/5, 5/5, 2/5 & 5/5 & 5/5 \\
1D linear elastic dynamics & 3/5 & 0/5 & 5/5 & 2/5 & 3/5 & 5/5 & 0/5 & 1/5 & 2/5 & 1/5, 1/5, 2/5 & 5/5, 5/5, 5/5 & 3/5, 2/5, 3/5 & 2/5, 5/5, 3/5 & 0/5, 1/5, 3/5 & 0/5, 5/5, 1/5 & 5/5 & 5/5 \\
\midrule
2D circular mesh & 5/5 & 5/5 & 5/5 & 5/5 & 5/5 & 5/5 & 5/5 & 5/5 & 3/5 & 5/5, 5/5, 5/5 & 5/5, 5/5, 5/5 & 5/5, 5/5, 5/5 & 5/5, 5/5, 5/5 & 4/5, 5/5, 5/5 & 4/5, 5/5, 5/5 & 5/5 & 5/5 \\
2D circular mesh (Delaunay) & 5/5 & 5/5 & 5/5 & 5/5 & 5/5 & 5/5 & 5/5 & 5/5 & 4/5 & 5/5, 5/5, 5/5 & 5/5, 5/5, 5/5 & 5/5, 5/5, 5/5 & 5/5, 5/5, 5/5 & 5/5, 5/5, 5/5 & 5/5, 5/5, 5/5 & 5/5 & 5/5 \\
2D Specht shape functions & 2/5 & 0/5 & 0/5 & 0/5 & 0/5 & 2/5 & 0/5 & 0/5 & 0/5 & 0/5, 2/5, 0/5 & 0/5, 0/5, 0/5 & 0/5, 0/5, 0/5 & 0/5, 0/5, 0/5 & 0/5, 0/5, 0/5 & 0/5, 0/5, 0/5 & 1/5 & 1/5 \\
2D Kirchhoff plate (T1) & 5/5 & 5/5 & 5/5 & 5/5 & 5/5 & 5/5 & 5/5 & 5/5 & 5/5 & 5/5, 5/5, 5/5 & 5/5, 5/5, 5/5 & 5/5, 5/5, 5/5 & 5/5, 5/5, 5/5 & 5/5, 5/5, 5/5 & 3/5, 5/5, 5/5 & 5/5 & 5/5 \\
2D Kirchhoff plate (T2) & 5/5 & 4/5 & 5/5 & 5/5 & 5/5 & 5/5 & 2/5 & 1/5 & 3/5 & 5/5, 5/5, 5/5 & 4/5, 5/5, 5/5 & 2/5, 2/5, 3/5 & 2/5, 5/5, 2/5 & 5/5, 4/5, 4/5 & 0/5, 0/5, 3/5 & 5/5 & 5/5 \\
2D Kirchhoff plate (T3) & 0/5 & 0/5 & 3/5 & 0/5 & 0/5 & 1/5 & 0/5 & 0/5 & 0/5 & 2/5, 3/5, 1/5 & 0/5, 0/5, 0/5 & 0/5, 0/5, 0/5 & 0/5, 0/5, 0/5 & 0/5, 0/5, 0/5 & 0/5, 0/5, 0/5 & 0/5 & 3/5 \\
2D hyperelasticity (T1) & 5/5 & 5/5 & 4/5 & 5/5 & 5/5 & 5/5 & 5/5 & 5/5 & 5/5 & 5/5, 5/5, 5/5 & 5/5, 5/5, 5/5 & 5/5, 5/5, 5/5 & 5/5, 5/5, 5/5 & 4/5, 5/5, 5/5 & 4/5, 5/5, 3/5 & 5/5 & 5/5 \\
2D hyperelasticity (T3) & 4/5 & 3/5 & 2/5 & 5/5 & 5/5 & 5/5 & 0/5 & 1/5 & 1/5 & 2/5, 5/5, 5/5 & 2/5, 4/5, 4/5 & 0/5, 1/5, 3/5 & 1/5, 3/5, 1/5 & 1/5, 2/5, 0/5 & 0/5, 2/5, 0/5 & 5/5 & 5/5 \\
2D cohesive zone (T1) & 5/5 & 5/5 & 5/5 & 5/5 & 4/5 & 3/5 & 5/5 & 3/5 & 0/5 & 5/5, 5/5, 4/5 & 5/5, 5/5, 5/5 & 4/5, 5/5, 3/5 & 4/5, 5/5, 3/5 & 4/5, 2/5, 0/5 & 2/5, 5/5, 5/5 & 4/5 & 5/5 \\
2D cohesive zone (T2) & 5/5 & 5/5 & 3/5 & 3/5 & 5/5 & 2/5 & 1/5 & 1/5 & 0/5 & 4/5, 2/5, 5/5 & 2/5, 3/5, 4/5 & 1/5, 2/5, 1/5 & 5/5, 4/5, 0/5 & 0/5, 3/5, 1/5 & 0/5, 2/5, 1/5 & 5/5 & 5/5 \\
2D cohesive zone (T3) & 5/5 & 1/5 & 3/5 & 3/5 & 2/5 & 5/5 & 0/5 & 1/5 & 0/5 & 5/5, 5/5, 5/5 & 4/5, 4/5, 4/5 & 1/5, 2/5, 1/5 & 2/5, 1/5, 0/5 & 1/5, 0/5, 1/5 & 0/5, 2/5, 1/5 & 5/5 & 5/5 \\
\midrule
3D hex8 linear elasticity & 5/5 & 5/5 & 5/5 & 5/5 & 5/5 & 5/5 & 5/5 & 2/5 & 0/5 & 4/5, 5/5, 5/5 & 3/5, 5/5, 5/5 & 3/5, 3/5, 5/5 & 4/5, 5/5, 4/5 & 5/5, 5/5, 4/5 & 1/5, 5/5, 1/5 & 5/5 & 5/5 \\
3D linear elastic dynamics (T1) & 5/5 & 5/5 & 5/5 & 5/5 & 5/5 & 5/5 & 5/5 & 3/5 & 2/5 & 5/5, 5/5, 5/5 & 5/5, 5/5, 5/5 & 5/5, 4/5, 0/5 & 5/5, 4/5, 4/5 & 5/5, 3/5, 4/5 & 2/5, 3/5, 1/5 & 5/5 & 5/5 \\
3D linear elastic dynamics (T3) & 5/5 & 2/5 & 5/5 & 5/5 & 3/5 & 5/5 & 1/5 & 1/5 & 0/5 & 4/5, 5/5, 5/5 & 1/5, 2/5, 5/5 & 0/5, 1/5, 2/5 & 1/5, 2/5, 3/5 & 3/5, 5/5, 2/5 & 0/5, 2/5, 1/5 & 5/5 & 5/5 \\
\midrule
\textbf{Function implementation (\%)} & 87.1 & 68.2 & 82.4 & 80.0 & 78.8 & 82.4 & 52.9 & 51.8 & 40.0 & 78.8, 85.9, 84.7 & 68.2, 75.3, 82.4 & 54.1, 57.6, 56.5 & 62.4, 72.9, 52.9 & 57.6, 60.0, 57.6 & 32.9, 65.9, 44.7 & 88.2 & \textbf{92.9} \\
\textbf{Pass@3 (Implementation) (\%)} & 93.5 & 79.4 & 93.5 & 87.6 & 87.6 & 95.9 & 62.9 & 73.5 & 55.3 & 90.6, 96.5, 91.2 & 84.1, 85.3, 88.2 & 70.6, 80.6, 76.5 & 81.8, 85.3, 72.9 & 71.2, 78.2, 71.2 & 49.4, 80.0, 67.6 & 91.8 & \textbf{97.6} \\
\textbf{Pass$^3$ (Implementation) (\%)} & 79.4 & 54.7 & 68.8 & 71.8 & 68.2 & 71.2 & 47.1 & 36.5 & 23.5 & 65.9, 77.1, 78.8 & 52.4, 64.1, 72.4 & 38.8, 38.2, 37.6 & 42.9, 57.1, 30.6 & 42.9, 41.2, 42.9 & 14.1, 53.5, 27.1 & 84.7 & \textbf{88.8} \\
\bottomrule
\end{tabular}%
}
\end{sidewaystable}

\begin{sidewaystable}[p]
\centering
\caption{\textbf{Single-pass test-generation performance by task.} Entries report joint test success rate (JTS, \%), averaged over five trials per task using Equation~\ref{eq:joint_test_success}. The final row averages equally across all 17 tasks. Bold marks the highest task-averaged score.}
\label{tab:task_model_test_generation_jts}
\fontsize{8}{10}\selectfont
\renewcommand{\arraystretch}{2.0}
\setlength{\tabcolsep}{3pt}

\resizebox{\linewidth}{!}{%
\begin{tabular}{@{}l*{15}{c}@{}}
\toprule
\textbf{Task} & \shortstack{Claude\\Fable 5.1} & \shortstack{Claude\\Sonnet 5} & \shortstack{GPT-6\\Astra} & \shortstack{Gemini\\3.1 Pro\\Preview} & \shortstack{Gemini\\3.7 Flash} & \shortstack{Gemini\\4 Argon} & \shortstack{Gemma\\4 31B} & \shortstack{DeepSeek\\V4 Pro\\0813} & \shortstack{Muse\\Glimmer\\30B} & \shortstack{Claude\\Opus 5} & \shortstack{GPT-5.6\\Sol} & \shortstack{GPT-5.6\\Luna} & \shortstack{Kimi K3} & \shortstack{GLM-5.2} & \shortstack{DeepSeek\\V4 Flash\\0731} \\
\midrule
1D isotropic hardening & 40.0 & 60.0 & 100.0 & 26.7 & 70.0 & 100.0 & 40.0 & 66.7 & 83.3 & 53.3 & 93.3 & 86.7 & 63.3 & 83.3 & 43.3 \\
1D kinematic hardening & 46.7 & 66.7 & 100.0 & 70.0 & 90.0 & 96.7 & 73.3 & 83.3 & 83.3 & 86.7 & 96.7 & 93.3 & 80.0 & 83.3 & 70.0 \\
1D linear elastic dynamics & 100.0 & 86.7 & 100.0 & 93.3 & 93.3 & 100.0 & 80.0 & 86.7 & 100.0 & 93.3 & 100.0 & 93.3 & 100.0 & 93.3 & 66.7 \\
\midrule
2D circular mesh & 100.0 & 80.0 & 86.7 & 80.0 & 93.3 & 100.0 & 80.0 & 93.3 & 40.0 & 100.0 & 86.7 & 86.7 & 100.0 & 93.3 & 53.3 \\
2D circular mesh (Delaunay) & 86.7 & 60.0 & 86.7 & 80.0 & 100.0 & 100.0 & 80.0 & 93.3 & 40.0 & 93.3 & 93.3 & 86.7 & 80.0 & 93.3 & 60.0 \\
2D Specht shape functions & 100.0 & 100.0 & 100.0 & 100.0 & 100.0 & 100.0 & 33.3 & 86.7 & 73.3 & 100.0 & 73.3 & 86.7 & 66.7 & 93.3 & 60.0 \\
2D Kirchhoff plate (T1) & 0.0 & 0.0 & 75.0 & 5.0 & 0.0 & 0.0 & 5.0 & 0.0 & 0.0 & 15.0 & 45.0 & 45.0 & 0.0 & 0.0 & 0.0 \\
2D Kirchhoff plate (T2) & 0.0 & 0.0 & 75.0 & 0.0 & 0.0 & 0.0 & 0.0 & 15.0 & 0.0 & 0.0 & 45.0 & 45.0 & 0.0 & 0.0 & 0.0 \\
2D Kirchhoff plate (T3) & 0.0 & 0.0 & 65.0 & 0.0 & 0.0 & 0.0 & 0.0 & 0.0 & 0.0 & 0.0 & 45.0 & 45.0 & 0.0 & 0.0 & 15.0 \\
2D hyperelasticity (T1) & 100.0 & 100.0 & 40.0 & 100.0 & 100.0 & 80.0 & 60.0 & 40.0 & 100.0 & 100.0 & 80.0 & 100.0 & 20.0 & 80.0 & 60.0 \\
2D hyperelasticity (T3) & 100.0 & 100.0 & 60.0 & 100.0 & 100.0 & 100.0 & 40.0 & 80.0 & 80.0 & 100.0 & 100.0 & 100.0 & 80.0 & 40.0 & 80.0 \\
2D cohesive zone (T1) & 16.0 & 0.0 & 96.0 & 68.0 & 32.0 & 8.0 & 12.0 & 15.0 & 24.0 & 56.0 & 68.0 & 72.0 & 16.0 & 32.0 & 20.0 \\
2D cohesive zone (T2) & 0.0 & 0.0 & 92.0 & 68.0 & 40.0 & 0.0 & 12.0 & 27.0 & 40.0 & 56.0 & 68.0 & 52.0 & 0.0 & 24.0 & 16.0 \\
2D cohesive zone (T3) & 32.0 & 8.0 & 76.0 & 60.0 & 44.0 & 24.0 & 8.0 & 14.0 & 24.0 & 72.0 & 68.0 & 60.0 & 24.0 & 52.0 & 20.0 \\
\midrule
3D hex8 linear elasticity & 26.7 & 66.7 & 60.0 & 86.7 & 66.7 & 40.0 & 60.0 & 53.3 & 60.0 & 53.3 & 60.0 & 26.7 & 60.0 & 60.0 & 40.0 \\
3D linear elastic dynamics (T1) & 0.0 & 0.0 & 92.0 & 36.0 & 0.0 & 36.0 & 0.0 & 0.0 & 0.0 & 40.0 & 84.0 & 84.0 & 16.0 & 0.0 & 16.0 \\
3D linear elastic dynamics (T3) & 100.0 & 0.0 & 100.0 & 13.3 & 0.0 & 20.0 & 6.7 & 0.0 & 60.0 & 80.0 & 93.3 & 100.0 & 20.0 & 0.0 & 0.0 \\
\midrule
\textbf{Mean JTS (\%)} & 49.9 & 42.8 & \textbf{82.6} & 58.1 & 54.7 & 53.2 & 34.7 & 44.4 & 47.5 & 64.6 & 76.5 & 74.3 & 42.7 & 48.7 & 36.5 \\
\bottomrule
\end{tabular}%
}
\end{sidewaystable}

\begin{table}[]
\centering
\caption{\textbf{Self-verification failure categories by model.} Counts pool failed SimpleAgent and OpenHands submissions. Cases gives the number of analyzed failed submissions for each model. Final labels combine 104 unanimous three-judge decisions and 145 adjudicated decisions. A case may receive multiple categories, so category counts can exceed the number of cases. Codes are expanded in Table~\ref{tab:failure_category_distribution}.}
\label{tab:failure_categories_by_model}
\small
\renewcommand{\arraystretch}{1.1}
\setlength{\tabcolsep}{4pt}

\begin{tabular}{@{}l c c c c c c c c c@{}}
\toprule
\textbf{Model} & \textbf{Cases} & \textbf{VF-01} & \textbf{VF-02} & \textbf{VF-03} & \textbf{VF-04} & \textbf{VF-05} & \textbf{VF-06} & \textbf{NV} & \textbf{Other} \\
\midrule
Claude Opus 5 & 19 & 10 & 13 & 0 & 7 & 1 & 0 & 0 & 0 \\
GPT-5.6 Sol & 26 & 16 & 17 & 0 & 7 & 0 & 1 & 0 & 0 \\
GPT-5.6 Luna & 68 & 59 & 55 & 2 & 11 & 4 & 10 & 0 & 0 \\
Kimi K3 & 37 & 25 & 21 & 4 & 16 & 10 & 2 & 0 & 0 \\
GLM-5.2 & 40 & 29 & 23 & 4 & 13 & 6 & 1 & 0 & 0 \\
DeepSeek V4 Flash 0731 & 59 & 43 & 46 & 1 & 17 & 9 & 1 & 0 & 0 \\
\midrule
\textbf{All models} & 249 & 182 & 175 & 11 & 71 & 30 & 15 & 0 & 0 \\
\bottomrule
\end{tabular}
\end{table}

\begin{table}[]
\centering
\caption{\textbf{Distribution of self-verification failure categories.} Counts cover 146 SimpleAgent and 103 OpenHands failed submissions. Percentages use all 249 analyzed failed cases as the denominator. A case may receive multiple labels, so category percentages need not sum to 100\%. The last row reports cohort sizes rather than a sum of category counts.}
\label{tab:failure_category_distribution}
\small
\renewcommand{\arraystretch}{1.1}
\setlength{\tabcolsep}{4pt}

\begin{tabular}{@{}l l c c c c@{}}
\toprule
\textbf{Code} & \textbf{Category} & \textbf{SimpleAgent} & \textbf{OpenHands} & \textbf{All} & \textbf{All (\%)} \\
\midrule
VF-01 & Inadequate Test Configuration and Coverage & 105 & 77 & 182 & 73.1 \\
VF-02 & Generic Consistency Checks & 103 & 72 & 175 & 70.3 \\
VF-04 & Non-Independent Verification & 40 & 31 & 71 & 28.5 \\
VF-05 & Test Accommodation & 17 & 13 & 30 & 12.0 \\
VF-06 & Unverified Post-Test Modification & 14 & 1 & 15 & 6.0 \\
VF-03 & Ineffective Test Assertions & 9 & 2 & 11 & 4.4 \\
NV & No Verification & 0 & 0 & 0 & 0.0 \\
Other & Other Verification Failure & 0 & 0 & 0 & 0.0 \\
\midrule
 & \textbf{Analyzed cases} & 146 & 103 & 249 & 100.0 \\
\bottomrule
\end{tabular}
\end{table}

\clearpage
\section{Task structure}
\label{app:task_prompt_templates}

\subsection{Task structure}
Each task is stored as a Python module containing the target function, any supplied helpers, reference tests, designated incorrect implementations, and task metadata. Listing~\ref{lst:task_structure} shows how these components are organized. The target signature and docstring define the requested behavior, including input and output conventions and numerical requirements. The reference function body, verification inputs, reference test bodies, and incorrect implementations are used for evaluation and are excluded from the task prompts.

\begin{lstlisting}[style=tasktemplate,caption={Schematic task module. Docstring fields are included where relevant to the task.},label={lst:task_structure}]
def target_function(<parameters and default values>):
    """
    <Physical problem, numerical formulation, and required computation.>

    Parameters
    ----------
    <name> : <type and array shape>
        <Meaning, units, indexing, and valid values.>

    Returns
    -------
    <name> : <type and array shape>
        <Meaning, ordering, and time/load-step convention.>

    Raises
    ------
    <exception type> : <condition requiring the exception>

    Notes
    -----
    <Constitutive equations, discretization, solver requirements,
     boundary conditions, or other task-specific conventions.>
    """
    ...  # Reference implementation body withheld from the model.

def test_required_behavior(fcn):
    """<Behavior to verify and any prescribed test configuration.>"""
    ...  # Reference test body withheld from the model.

def task_info():
    return {
        "task_id": "<task identifier>",
        "task_short_description": "<brief description>",
        "main_fcn": target_function,
        "required_imports": [<permitted import statements>],
        "fcn_dependencies": [<supplied helper functions>],
        "reference_verification_inputs": [<input argument sets>],
        "test_cases": [{
            "test_code": test_required_behavior,
            "expected_failures": [<incorrect implementations>],
        }],
    }
\end{lstlisting}

\clearpage
\subsection{Example task specification}
Listing~\ref{lst:task_example} shows the signature and selected docstring passages from the three-dimensional dynamics task. The specification fixes array shapes, node and degree-of-freedom ordering, loading, integration parameters, and the interpretation of the returned displacement history.

\begin{lstlisting}[style=tasktemplate,caption={Excerpt from the three-dimensional linear elastic dynamics task specification.},label={lst:task_example}]
def FEM_3D_linelast_dynamic_CC1_H5_T1(
    coords: np.ndarray, connect: np.ndarray,
    fixed_node_indices: np.ndarray, F_ext: np.ndarray,
    E: float, nu: float, rho: float, dt: float,
    total_steps: int, beta1: float = 0.5, beta2: float = 0.0,
) -> np.ndarray:
    """
    Solve a 3D linear-elastic structural dynamics problem on a hexahedral mesh
    using the Newmark-Beta time integration method.
    [...]

    Parameters
    ----------
    coords : (N, 3) ndarray of float
        Global coordinates [x, y, z] for each of the N nodes.
    connect : (E, 8) ndarray of int
        Element connectivity for 8-node hexahedral (Hex8) elements. Each row
        contains the 0-based node indices of one element.
    fixed_node_indices : (K,) ndarray of int
        Indices of nodes that are fully clamped (all 3 translational DOFs fixed).
    F_ext : (3*N,) ndarray of float
        Global external force vector applied as a constant step load at every
        time step. DOF ordering: [u_x0, u_y0, u_z0, u_x1, ...].
    [...]
    beta1 : float, optional
        Newmark parameter controlling velocity update (artificial damping).
        Default is 0.5 (no numerical damping).
    beta2 : float, optional
        Newmark parameter controlling displacement update (integration scheme).
        Default is 0.0 (explicit central-difference method).

    Returns
    -------
    history : (total_steps, 3*N) ndarray of float
        Displacement of every DOF at every time step. Row i contains the
        displacement vector at time (i+1)*dt.

    Notes
    -----
    - 3 DOFs per node: [u_x, u_y, u_z].
    - The mass matrix is lumped (diagonal) using the row-sum technique.
    [...]
    """
    ...
\end{lstlisting}

\clearpage
\clearpage
\section{Agent harness configurations}
\label{app:harness_details}

This appendix documents the configurations and execution protocols for each agentic harness. The full prompts, including controller messages returned after executions or rejected actions, are provided in the study repository.

\lstdefinestyle{harnessprompt}{
  style=tasktemplate,
  basicstyle=\ttfamily\footnotesize,
  breaklines=true,
  breakatwhitespace=false,
  literate={—}{{--}}1
}

\subsection*{Harness configurations}
All harnesses use a per-trial budget of 2,000,000 cumulative input and output tokens, a 7,200-second wall-time limit, and prompt caching. The token budget counts input and output across all model calls within the trial. For the coder-tester extension, the token and wall-time budgets are shared across both roles. Its iteration limit applies separately to each agent. 
No OpenHands run terminated because of the USD 75 cost ceiling.
Table~\ref{tab:harness_limits} gives the remaining limits.

\begin{table}[!htbp]
\centering
\caption{\textbf{Agent harness operational limits.} Timeouts apply separately to each action or request. Other limits apply per trial unless stated. A dash indicates that no explicit setting is reported here.}
\label{tab:harness_limits}
\label{tab:simpleagent_limits}
\label{tab:openhands_limits}
\label{tab:coder_tester_limits}
\small
\renewcommand{\arraystretch}{1.15}
\setlength{\tabcolsep}{4pt}
\begin{tabularx}{\linewidth}{@{}>{\raggedright\arraybackslash}p{0.30\linewidth} *{3}{>{\raggedright\arraybackslash}X}@{}}
\toprule
Setting & SimpleAgent & OpenHands & coder-tester \\
\midrule
Output tokens per call & 30,000 & 30,000 & 32,000 (either role) \\
Iterations / steps & 200 & 200 & 200 per agent \\
Python / scratch timeout & 120 s & -- & 60 s \\
Full-suite timeout & -- & -- & 300 s \\
Execution-output limit & 10,000 characters & -- & 10,000 characters \\
Coder scratch executions & -- & -- & 3 per candidate \\
Candidate versions & -- & -- & 50 \\
Model-request timeout & -- & 600 s & -- \\
\texttt{model\_retry\_attempts} & -- & 1 & -- \\
API-cost ceiling & -- & USD 75 per task & -- \\
Network access & Disabled & Enabled & Disabled \\
\bottomrule
\end{tabularx}
\end{table}

\clearpage
\section{LLM-as-judge pipeline configuration details}

\subsection{Judge configuration}
\label{app:judge_configuration}
Table~\ref{tab:judge_configuration} lists the model configuration used for the judge models. The three initial judges assess each case independently. Cases with disagreements in any label are sent to GPT-6 Astra for adjudication. Each request contains the prepared case package and the corresponding judging instructions. No explicit temperature or sampling seed is configured. Requests that fail or return an invalid judgment are retried, with up to three attempts per invocation. One valid classification is retained for each case and judge.

\begin{table}[!htbp]
\centering
\caption{\textbf{LLM judge configuration.} Output token caps are measured in tokens per call. High denotes the configured reasoning effort. Claude Opus 5.5 uses adaptive reasoning with high effort.}
\label{tab:judge_configuration}
\small
\renewcommand{\arraystretch}{1.2}
\setlength{\tabcolsep}{3pt}
\begin{tabular}{@{}llllr@{}}
\toprule
\textbf{Model} & \textbf{Role} & \textbf{Provider} & \textbf{Reasoning} & \textbf{Output token cap} \\
\midrule
GPT-6.1 Sol & Initial judge & AWS & High & 131,072 \\
Claude Opus 5.5 & Initial judge & AWS & Adaptive (high) & 128,000 \\
Grok 4.7 & Initial judge & AWS & High & 128,000 \\
GPT-6 Astra & Adjudicator & AWS & High & 128,000 \\
\bottomrule
\end{tabular}
\end{table}

\subsection{Agreement among the initial judges}
\label{app:judge_agreement}
We measure agreement using the three initial judgments for each of the $N=249$ completed failed submissions, before adjudication. Because the classification is multi-label, each category is treated as a separate binary decision indicating presence or absence. Fleiss' $\kappa$~\citep{fleiss1971agreement} adjusts the observed agreement for the agreement expected by chance. The expected agreement is calculated from the pooled frequencies of category presence and absence.

For a given category, let $\bar{P}$ denote mean pairwise agreement, and $P_e$ denote the expected agreement. Then the chance-adjusted Fleiss' $\kappa$ is 
\begin{equation}
\label{eq:fleiss_kappa}
\kappa=\frac{\bar{P}-P_e}{1-P_e}.
\end{equation}
Here, $\kappa=1$ indicates complete agreement. 
Table~\ref{tab:judge_agreement} reports Fleiss' $\kappa$ and the prevalence of each category.

\begin{table}[!htbp]
\centering
\caption{\textbf{Initial-judge agreement and category prevalence.} Fleiss' $\kappa$ measures agreement among three judges across 249 failed submissions before adjudication. Prevalence is the percentage of the 747 individual judgments assigning each category.}
\label{tab:judge_agreement}
\small
\renewcommand{\arraystretch}{1.2}
\setlength{\tabcolsep}{4pt}
\begin{tabular}{@{}llcc@{}}
\toprule
\textbf{Code} & \textbf{Category} & \textbf{Prevalence (\%)} & \textbf{$\kappa$} \\
\midrule
VF-01 & Inadequate Test Configuration and Coverage & 68.1 & 0.612 \\
VF-02 & Generic Consistency Checks & 66.1 & 0.665 \\
VF-03 & Ineffective Test Assertions & 3.1 & 0.551 \\
VF-04 & Non-Independent Verification & 23.4 & 0.731 \\
VF-05 & Test Accommodation & 10.0 & 0.718 \\
VF-06 & Unverified Post-Test Modification & 4.7 & 0.880 \\
\bottomrule
\end{tabular}
\end{table}

\end{document}